\documentclass[prd,twocolumn,showpacs,floatfix,amsmath,amssymb,nofootinbib]{revtex4}
\usepackage{bm}
\usepackage{graphicx}
\newcommand{\ud}{\mathrm{d}}

\newcommand{\GF}{G_\mathrm{F}}
\newcommand{\Hb}{\mathbf{H}}

\newcommand{\rX}{r_\mathrm{X}}
\newcommand{\Sb}{\mathbf{S}}
\newcommand{\sB}{\mathbf{s}}
\newcommand{\eff}{\mathrm{eff}}
\newcommand{\HV}{\Hb_\mathrm{V}}
\newcommand{\EC}{E_\mathrm{C}}
\newcommand{\muV}[1]{\mu_{\mathrm{V} #1}}
\newcommand{\qb}{\mathbf{q}}
\newcommand{\rb}{\mathbf{r}}
\newcommand{\Qb}{\mathbf{Q}}
\newcommand{\Jb}{\mathbf{J}}
\newcommand{\rB}{\mathbf{r}}
\newcommand{\gB}{\mathbf{g}}
\newcommand{\wsync}{\Omega_\mathrm{sync}}
\newcommand{\basef}[1]{\bm{\hat{\mathbf{e}}}^\mathrm{f}_{#1}}

\newcommand{\basev}[1]{\bm{\hat{\mathbf{e}}}^\mathrm{v}_{#1}}
\newcommand{\thetav}{\theta_\mathrm{v}}
\newcommand{\qthetav}{\tilde{\theta}_\mathrm{v}}
\newcommand{\nnuc}{n_\nu^\mathrm{c}}
\newcommand{\nnux}{n_\nu^\mathrm{x}}
\newcommand{\nratio}{\frac{n_\nu}{n_\nu^0}}

\newcommand{\myfigsep}{0.04 \textwidth}
\newcommand{\myfigwid}{0.48 \textwidth}

\begin{document}
\title{Analysis of Collective Neutrino Flavor Transformation in Supernovae}
\newcommand*{\UCSD}{Department of Physics, %
University of California, San Diego, %
La Jolla, CA 92093-0319}
\affiliation{\UCSD}
\newcommand*{\LANL}{Theoretical Division, Los Alamos National Laboratory, %
Los Alamos, NM 87545}
\affiliation{\LANL}
\newcommand*{\UMN}{School of Physics and Astronomy, %
University of Minnesota, Minneapolis, MN 55455}
\affiliation{\UMN}

\author{Huaiyu Duan}
\email{hduan@ucsd.edu}
\affiliation{\UCSD}
\author{George M.~Fuller}
\email{gfuller@ucsd.edu}
\affiliation{\UCSD}
\author{J.~Carlson}
\email{carlson@lanl.gov}
\affiliation{\LANL}
\author{Yong-Zhong Qian}
\email{qian@physics.umn.edu}
\affiliation{\UMN}

\date{\today}

\begin{abstract}
We study the flavor evolution of a dense gas initially consisting
of pure mono-energetic $\nu_e$ and $\bar\nu_e$. Using 
adiabatic invariants and the special symmetry in such a system
we are able to calculate the flavor evolution of the neutrino
gas for the cases with slowly decreasing neutrino 
number densities. These calculations give new insights
into the results of recent
large-scale numerical simulations of neutrino flavor
transformation in supernovae. For example, 
our calculations reveal the existence of what we
term the ``collective precession mode''. Our analyses
suggest that neutrinos which travel on intersecting
trajectories subject to destructive quantum interference
nevertheless can be in this mode. This  mode can result in
sharp transitions in the final
energy-dependent neutrino survival probabilities across
all trajectories, a feature seen in the numerical simulations.
Moreover, this transition is qualitatively different for the
normal and inverted neutrino mass hierarchies. Exploiting  this
difference, the neutrino signals from a future galactic supernova
can potentially be used to determine the actual
neutrino mass hierarchy.
\end{abstract}


\maketitle

\section{Introduction}

In this paper we employ physical, analytic insights
along with the results of large-scale numerical
calculations to study the nature of collective neutrino
and antineutrino flavor transformation in supernovae.
Although neutrino flavor transformation is an experimental
fact, modeling this process in astrophysical settings
can be problematic. In part, this is because nature
produces environments where the number densities of
neutrinos and/or antineutrinos can be very large.
Examples of these include the early universe and
environments associated with compact-object mergers
and gravitational collapse. In particular, core-collapse supernovae
result in hot proto-neutron stars that
 emit neutrinos and antineutrinos copiously from
the neutrino sphere. This implies inhomogeneous,
anisotropic distributions for these particles.
As their trajectories intersect above the proto-neutron star,
their flavor evolution histories 
are quantum mechanically coupled \cite{Qian:1994wh}.
The flavor content of the neutrino and antineutrino fields 
in and above a proto-neutron star will be a necessary
ingredient for the interpretation of neutrino signals
from a future supernova. It can also be an important, even crucial
determinant of the composition of supernova ejecta \cite{Qian:1993dg} and
possibly even the supernova explosion mechanism \cite{Fuller:1992aa}.
Consequently, if we are to understand core-collapse supernovae, it follows
that we must understand neutrino and antineutrino flavor
evolution in them.

Large neutrino number densities imply that neutrino-neutrino
in addition to neutrino-electron forward scattering sets
the potential which governs neutrino flavor conversion.
Because of the neutrino-neutrino forward scattering potential
\cite{Fuller:1987aa,Pantaleone:1992xh,Sigl:1992fn}, 
neutrino flavor transformation
in the early universe and near the supernova core can be
very different from that in the vacuum or in an ordinary matter background only
\cite{Samuel:1993uw,Kostelecky:1993ys,%
Kostelecky:1993yt,Kostelecky:1993dm,Kostelecky:1994dt,%
Samuel:1995ri,Kostelecky:1996bs,Pastor:2001iu,Dolgov:2002ab,%
Wong:2002fa,Abazajian:2002qx}.
Recent analytical and numerical studies have revealed a new paradigm for
neutrino flavor transformation in supernovae
\cite{Pastor:2002we,Balantekin:2004ug,Fuller:2005ae,%
Duan:2005cp,Duan:2006an,Duan:2006jv,Hannestad:2006nj,%
Balantekin:2006tg,Raffelt:2007yz},
one which is completely different from vacuum oscillations 
or the conventional Mikheyev-Smirnov-Wolfenstein (MSW) mechanism
\cite{Wolfenstein:1977ue,Wolfenstein:1979ni,Mikheyev:1985aa}.

A particular aspect of this 
new paradigm is best discussed in the following framework:
The $2\times2$ neutrino flavor transformation problem
can be described as the motion of isospins in  flavor space,
wherein $\nu_e/\bar\nu_{\mu,\tau}$ and $\bar\nu_e/\nu_{\mu,\tau}$ correspond
to the ``up'' and ``down'' states of these isospins \cite{Duan:2005cp}.
Both analytical and numerical studies have suggested that
 a dense neutrino gas originally in 
a ``bipolar configuration'' 
(\textit{i.e.}, with the corresponding 
neutrino flavor isospins or NFIS's forming two oppositely
oriented groups) tends to stay in such a  configuration
even though each isospin group is composed of neutrinos and/or
antineutrinos with finite energy spread.
In other words, neutrinos and antineutrinos with different energies
can experience collective flavor transformation 
at  high neutrino number densities.
This is very different from the conventional MSW paradigm in which
neutrinos and antineutrinos with different energies undergo flavor
transformation independently.

A neutrino system with a bipolar configuration 
is also referred to as a ``bipolar system''.
Because supernova neutrinos are essentially
in their flavor eigenstates when they leave the neutrino sphere,
they naturally form a bipolar system.
The neutrino sphere is in the very high density,
electron degenerate environment near the
neutron star surface.

For a simple bipolar system consisting of mono-energetic
$\nu_e$ and $\bar\nu_e$ initially,
it has been shown that the evolution of the system is
equivalent to the motion of a (gyroscopic) pendulum \cite{Hannestad:2006nj}.
Therefore, a bipolar system generally can evolve simultaneously
in two different kinds of modes, \textit{i.e.} the precession mode
and the nutation mode, in analogy to the mechanical motion of
a gyroscopic pendulum. In the extreme limit of large neutrino number density,
a bipolar system is reduced to a synchronized system, which
is in a pure precession mode characterized by
a common synchronization frequency \cite{Pastor:2001iu}. 
The evolution of bipolar
systems in the presence of an ordinary matter background has been studied
in Refs.~\cite{Duan:2005cp,Hannestad:2006nj} using 
corotating frames.

Refs.~\cite{Duan:2006an,Duan:2006jv} have presented by far the most
sophisticated, large-scale numerical simulations of neutrino
flavor transformation in the coherent regime near the
supernova core. For example, these simulations
for the first time  self-consistently treated the evolution of
neutrinos propagating along various intersecting trajectories.
These simulations clearly show that the conventional
MSW paradigm is invalid near the supernova core where
neutrino fluxes are large. However, analytical models so far
have only corroborated some of the features demonstrated by
the simulations, and there are some obvious gaps between the
analytical and numerical studies.

One of the gaps is that current analytical models of bipolar
systems assume constant neutrino number densities,
which is not true in supernovae. Using some simple numerical
examples, Ref.~\cite{Hannestad:2006nj} has shown that
some interesting phenomena observed in the  simulations 
\cite{Duan:2006an,Duan:2006jv}
are related to varying neutrino number densities. For example, the
energy averaged neutrino survival probabilities change
as neutrino number densities decrease with the radius. In this paper
we will show that if the neutrino number density decreases slowly
as the system  evolves out of the synchronized limit
at high neutrino densities,
the bipolar system will be dominantly in a precession mode. 
The neutrino flavor evolution  seen in the numerical
simulations is the combined effect of this precession mode and
the nutation modes that are generated as a result of the 
finite rate of change in neutrino number densities.

Another important gap between analytical and numerical studies
is that most of the current analytical models
assume homogeneity and isotropy of the neutrino gas, which is 
not true of the supernova environment. 
A recent analytical study which assumes an initial state of
$\nu_e$ and $\bar\nu_e$ with equal densities
shows that the collectivity (referred to as
``coherence'' or ``kinematic coherence'' in Ref.~\cite{Hannestad:2006nj})
of the nutation modes tends to break down quickly among different
neutrino trajectories in an
inhomogeneous and anisotropic environment
\cite{Raffelt:2007yz}. However, the numerical simulations
presented in Refs.~\cite{Duan:2006an,Duan:2006jv}
employed more realistic supernova conditions
where the initial $\nu_e$ and $\bar\nu_e$ as well as
$\nu_{\mu,\tau}$ and $\bar\nu_{\mu,\tau}$ do not have
the same number densities. These simulations do show
some clear signs of collective flavor transformation.
One important example is the hallmark pattern in the final
energy-dependent neutrino survival probability which has a sharp transition
energy across all neutrino trajectories (Fig.~3 of Ref.~\cite{Duan:2006jv}).
We have further analyzed the numerical results obtained
in the large-scale simulations mentioned above and found that,
apart from the non-collective nutation modes,
neutrinos propagating along different trajectories were in a single,
collective precession mode. It is this precession mode that facilitates
the mechanism suggested in Ref.~\cite{Duan:2006an}
for producing the fore-mentioned hallmark pattern in the final neutrino 
survival probability.

This paper is organized as follows.
In Sec.~\ref{sec:symmetric-sys} we will study the properties
of a symmetric bipolar
system initially consisting of an equal number of $\nu_e$ and $\bar\nu_e$.
We will use an adiabatic invariant of the system to obtain
some analytical understanding of the evolution of such a system
as  neutrino number densities change.
In Sec.~\ref{sec:asymmetric-sys} we will 
compare a simple asymmetric bipolar system with a gyroscopic
pendulum. We will revisit the criteria determining whether a bipolar system
is in the synchronized or bipolar regime and clarify
the description of bipolar oscillations.
In Sec.~\ref{sec:precession-mode} we will show that an asymmetric bipolar
system can stay roughly in a pure precession mode if neutrino number
densities decrease slowly. We will also demonstrate some interesting
properties of such a precession mode which can explain the results
from the simple numerical examples of Ref.~\cite{Hannestad:2006nj}.
In Sec.~\ref{sec:supernovae} we will apply our simple analytical
models to understand the numerical simulations presented in 
Refs.~\cite{Duan:2006an,Duan:2006jv} and offer
some new analyses of these  simulations.
In Sec.~\ref{sec:conclusion} we give our conclusions.

\section{Symmetric Bipolar System%
\label{sec:symmetric-sys}}

\subsection{Flavor pendulum\label{sec:pendulum-eom}}
We start with a simple bipolar system initially consisting of
mono-energetic $\nu_e$ and $\bar\nu_e$ with energy $E_\nu$ and an equal number
density $n_\nu$. Throughout this paper we will assume $2\times2$ flavor mixing
through the active-active channel. 
According to Ref.~\cite{Duan:2005cp}, the flavor
evolution of a neutrino or antineutrino
 is equivalent to the
motion of the corresponding neutrino
flavor isospin, or NFIS,
in flavor space. For a neutrino, the NFIS in the flavor basis is defined as
\begin{equation}
\sB_\nu\equiv\frac{1}{2}
\begin{pmatrix}
2 \mathrm{Re} (a_{\nu_e}^* a_{\nu_\tau})\\
2 \mathrm{Im} (a_{\nu_e}^* a_{\nu_\tau}) \\
|a_{\nu_e}|^2-|a_{\nu_\tau}|^2
\end{pmatrix},
\end{equation}
where $a_{\nu_e}$ and $a_{\nu_\tau}$ are the amplitudes for the neutrino
to be  in $\nu_e$ and another flavor state, say $\nu_\tau$, respectively. For an
antineutrino, the corresponding NFIS in the flavor basis is 
\begin{equation}
\sB_{\bar\nu}\equiv -\frac{1}{2}
\begin{pmatrix}
2 \mathrm{Re} (a_{\bar\nu_e} a_{\bar\nu_\tau}^*)\\
2 \mathrm{Im} (a_{\bar\nu_e} a_{\bar\nu_\tau}^*) \\
|a_{\bar\nu_e}|^2-|a_{\bar\nu_\tau}|^2
\end{pmatrix},
\end{equation}
where $a_{\bar\nu_e}$ and $a_{\bar\nu_\tau}$ are the amplitudes for the 
antineutrino to be  $\bar\nu_e$ and  $\bar\nu_\tau$, respectively.

To obtain a simple analytical understanding of collective neutrino
flavor transformation, we will assume,  unless otherwise stated,
 that the neutrino gas is
homogeneous and isotropic and that there is no ordinary
matter medium.
Using the NFIS notation, the equations of motion (e.o.m.) for the NFIS's
$\sB_1$ (neutrino) and $\sB_2$ (antineutrino)
of this simple bipolar system are \cite{Duan:2005cp}
\begin{subequations}
\label{eq:nfis-eom}
\begin{align}
\frac{\ud}{\ud t} \sB_1 
&= \sB_1 \times (\muV{,1}\HV + \mu_\nu n_{\nu,2} \sB_2), 
\label{eq:s1-eom}\\
\frac{\ud}{\ud t} \sB_2 
&= \sB_2 \times (\muV{,2}\HV + \mu_\nu n_{\nu,1} \sB_1), 
\label{eq:s2-eom}
\end{align}
\end{subequations}
and the initial condition is
\begin{equation}
\sB_1(0) =-\sB_2(0) = \frac{\basef{z}}{2},
\label{eq:s12-ini}
\end{equation}
where 
\begin{equation}
n_{\nu,1} = n_{\nu,2} = n_\nu
\end{equation}
are the number densities of neutrinos and antineutrinos,
and $\basef{z}$ is the unit vector in the $z$-direction
in the flavor basis.
Eq.~\eqref{eq:nfis-eom} clearly shows that the motion of the NFIS's is
similar to that of magnetic spins. In this ``magnetic spin''
picture, the ``magnetic spins'' $\sB_1$ and $\sB_2$
 precess around a common ``magnetic field''
\begin{equation}
\HV \equiv -\basef{x}\sin 2\thetav + \basef{z}\cos 2\thetav
\end{equation}
 with  ``gyro magnetic ratios'' 
\begin{equation}
\muV{,1} = -\muV{,2} = \muV{} \equiv \frac{\delta m^2}{2 E_\nu}.
\end{equation}
At the same time, $\sB_1$ and $\sB_2$
are also coupled to each other with a  coefficient
\begin{equation}
\mu_\nu \equiv -2\sqrt{2}\GF,
\end{equation}
where $\GF$ is the Fermi constant.

In this paper we always take the squared difference
of the two neutrino vacuum mass eigenvalues to be positive
($\delta m^2 \equiv m_2^2-m_1^2>0$).
Accordingly, the vacuum mixing angle $\thetav$ varies within $(0,\pi/2)$.
A normal mass hierarchy corresponds to a mixing angle $\thetav$ with 
$0<\thetav<\pi/4$  and an inverted mass
hierarchy corresponds to $\pi/4<\thetav<\pi/2$.
For an inverted mass hierarchy scenario, we follow Ref.~\cite{Hannestad:2006nj}
to define 
\begin{equation}
\qthetav\equiv\frac{\pi}{2}-\thetav,
\end{equation}
 which has
$0<\qthetav<\pi/4$. We will loosely refer to both $\thetav$
and $\qthetav$ as vacuum mixing angles.%
\footnote{Refs.~\cite{Duan:2005cp,Duan:2006an,Duan:2006jv}
have adopted a different convention for the
inverted mass hierarchy scenario where $\qthetav$ defined here
\textit{is} the vacuum mixing angle and $\delta m^2$ is taken
to be negative.
We note that these two conventions
 are equivalent by the simultaneous transformations
$|\nu_1\rangle\leftrightarrow|\nu_2\rangle$ and 
$|\nu_\tau\rangle\rightarrow-|\nu_\tau\rangle$. Correspondingly, one has
$\HV\rightarrow-\HV$ and $\basef{x(y)}\rightarrow-\basef{x(y)}$
in flavor space. 
The $x$- and $y$-components of a NFIS in the flavor basis
under these two conventions
are different by a sign for the inverted mass hierarchy case.}

We first look at the scenario with $n_\nu$ being constant. 
With the definition of
\begin{equation}
\Sb_\pm \equiv n_{\nu,1}\sB_1 \pm n_{\nu,2}\sB_2,
\end{equation}
Eqs.~\eqref{eq:nfis-eom} and \eqref{eq:s12-ini} become \cite{Duan:2005cp}
\begin{subequations}
\label{eq:SpSm-eom}
\begin{align}
\frac{\ud}{\ud t} \Sb_+ 
&= \muV{} \Sb_- \times \HV, 
\label{eq:Sp-eom}\\
\frac{\ud}{\ud t} \Sb_-
&= \muV{} \Sb_+ \times \HV + \mu_\nu \Sb_-\times \Sb_+,
\label{eq:Sm-eom}
\end{align}
\end{subequations}
and 
\begin{subequations}
\label{eq:SpSm-ini}
\begin{align}
\Sb_+(0) &= 0,
\label{eq:Sp-ini}\\
\Sb_-(0) &= n_\nu \basef{z}.
\label{eq:Sm-ini}
\end{align}
\end{subequations}
It is more convenient to work in the vacuum mass basis where the unit
vectors $\basev{i}$ are related to those in the flavor basis $\basef{i}$ by
\begin{subequations}
\begin{align}
\basev{x} &= \basef{x}\cos 2\thetav + \basef{z}\sin 2\thetav,\\
\basev{y} &= \basef{y},\\
\basev{z} &= \HV = -\basef{x}\sin 2\thetav + \basef{z}\cos 2\thetav.
\end{align}
\end{subequations}
Using Eqs.~\eqref{eq:SpSm-eom} and \eqref{eq:SpSm-ini}
one can check explicitly that vector $\Sb_-$ rotates
in the $\basev{x}$-$\basev{z}$ plane while $\Sb_+$ varies only along the
$\basev{y}$-axis \cite{Duan:2005cp}. 

In reality neutrinos can experience collective oscillations
only if $n_\nu$ is large. The largeness of 
the neutrino number density in this simple bipolar system is naturally 
measured by the ratio $n_\nu/n_\nu^0$, where
\begin{equation}
n_\nu^0 \equiv \frac{\muV{}}{|\mu_\nu|}
=\frac{\delta m^2}{4\sqrt{2}\GF E_\nu}.
\end{equation} 
In the limit $n_\nu/n_\nu^0\gg1$, the last term
in Eq.~\eqref{eq:Sm-eom} dominates and
\begin{equation}
\label{eq:Sm-eom2}
\frac{\ud}{\ud t} \Sb_-
\simeq  \mu_\nu \Sb_-\times \Sb_+.
\tag{\ref{eq:Sm-eom}$^\prime$}
\end{equation}
Therefore, $\Sb_-$ roughly maintains a constant magnitude 
if the neutrino number density is large. As a result,
 $\sB_1$ and $\sB_2$ are always roughly anti-aligned 
although their directions can be completely overturned in some scenarios.
It is after this special property that ``bipolar'' flavor
transformation was initially named \cite{Duan:2005cp}.%
\footnote{Ref.~\cite{Hannestad:2006nj} appears to have misunderstood
the origin of the word ``bipolar'' by stating that
the notation ``bipolar oscillation''
is a ``misnomer''.}
 
We define $\vartheta$ as the angle
between vectors $\Sb_-$ and $\HV$, which varies
 within $(0,\pi)$
if $\Sb_-\cdot\basev{x}>0$ and  within $(-\pi,0)$
if $\Sb_-\cdot\basev{x}<0$. We also define
\begin{equation}
p_\vartheta \equiv \frac{\Sb_+\cdot\basev{y}}{n_\nu}.
\label{eq:p-def}
\end{equation}
With the initial condition in Eq.~\eqref{eq:SpSm-ini},
we find that
Eqs.~\eqref{eq:Sp-eom} and \eqref{eq:Sm-eom2} are equivalent to
\begin{subequations}
\label{eq:p-vartheta-eom}
\begin{align}
\dot{p}_\vartheta &= -\muV{}\sin\vartheta,\\
\dot\vartheta &\simeq \muV{}\left(\nratio\right) p_\vartheta.
\end{align}
\end{subequations}

\begin{figure}
\begin{center}
\includegraphics*[scale=0.35, keepaspectratio]{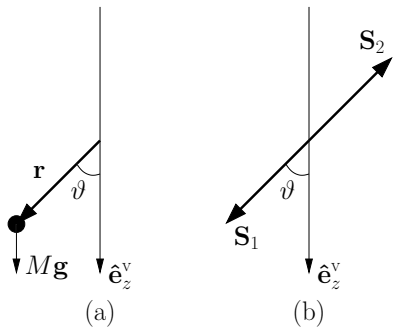}
\end{center}
\caption{\label{fig:pendulum}%
The equivalence of a pendulum and a symmetric 
($n_{\nu,1}=n_{\nu,2}=n_\nu$) bipolar system
initially consisting of pure $\nu_e$ and $\bar\nu_e$.
In the limit $n_\nu/n_\nu^0\gg1$ two NFIS blocks 
$\Sb_1=n_{\nu,1}\sB_1$ (for neutrino) and $\Sb_2=n_{\nu,2}\sB_2$ 
(for antineutrino) are always roughly
anti-aligned and $\Sb_1$ points roughly in the same direction
as the pendulum in flavor space.}
\end{figure}

Eq.~\eqref{eq:p-vartheta-eom} can be further reduced 
to a differential equation of
$\vartheta$ of the second order:
\begin{equation}
\ddot\vartheta\simeq-\omega^2\sin\vartheta,
\label{eq:vartheta-osc}
\end{equation}
where
\begin{subequations}
\begin{align}
\omega &\equiv \muV{}\sqrt{\nratio} \\
&=\left(\frac{\sqrt{2}\GF\delta m^2 n_\nu}{E_\nu}\right)^{1/2}
\end{align}
\end{subequations}
is an intrinsic frequency of the system.
Because Eq.~\eqref{eq:vartheta-osc} also describes 
the motion of a pendulum, we can view
the flavor transformation in this simple bipolar system
as the motion of a pendulum in the flavor
space (Fig.~\ref{fig:pendulum}).
We note that the mass $M$ of the ``flavor pendulum''
is irrelevant in this case. The only relevant parameter
is the ratio between the magnitude of the
acceleration field $\gB$ and the length of the 
pendulum $r$, which is  related to the intrinsic frequency by
\begin{equation}
\omega = \sqrt{\frac{g}{r}}.
\end{equation}
The period of the flavor pendulum is
(see, \textit{e.g.}, Ref.~\cite{Landau:1976aa})
\begin{equation}
\label{eq:T-pendulum}
T =\frac{4 \bm{K}(\sin(\vartheta_{\max}/2))}{\omega},
\end{equation}
where
\begin{equation}
\bm{K}(k) \equiv \int_0^{\pi/2} \frac{\ud \zeta}{\sqrt{1-k^2\sin^2\zeta}}
\end{equation}
is the complete elliptic integral of the first kind
\cite{Gradshteyn:1994aa}. 

The period of the simple symmetric bipolar system in Eq.~\eqref{eq:T-pendulum}
takes a simpler form if the vacuum mixing angle $\thetav$
or $\qthetav$ is small.
For the normal mass  hierarchy scenario with
$\thetav\ll 1$, the pendulum
motion is the same as that of a harmonic oscillator and
\begin{equation}
T \simeq \frac{2\pi}{\omega}
=2\pi\left(\frac{E_\nu}{\sqrt{2}\GF\delta m^2 n_\nu}\right)^{1/2}.
\label{eq:T-norm}
\end{equation}
For the inverted mass hierarchy  scenario with
$\qthetav\ll 1$, we expand Eq.~\eqref{eq:T-pendulum}
in terms of $\qthetav$ \cite{Gradshteyn:1994aa} and find that
\begin{equation}
T \simeq \frac{4\ln(4/\qthetav)}{\omega}
=4\ln(4/\qthetav)\left(\frac{E_\nu}{\sqrt{2}\GF\delta m^2 n_\nu}\right)^{1/2}.
\label{eq:T-inv}
\end{equation}
The period of the bipolar oscillation in this limit has a logarithmic 
dependence on $\tilde{\theta}_\mathrm{v}$ as pointed out in 
Ref.~\cite{Hannestad:2006nj}. The periods of bipolar oscillations
calculated using Eqs.~\eqref{eq:T-norm} and \eqref{eq:T-inv}
are in excellent agreement with the simple numerical examples
 in Ref.~\cite{Duan:2005cp}.

Ref.~\cite{Hannestad:2006nj}  has shown that the evolution
of this simple bipolar system is equivalent to a pendulum motion
for any $n_\nu$ (also see Sec.~\ref{sec:two-modes}). 
In the limit $n_\nu/ n_\nu^0\gg1$ the flavor pendulum
described here is the same as that in Ref.~\cite{Hannestad:2006nj}.
This limit is of interest to analyses of the flavor evolution
of supernova neutrinos and antineutrinos which have finite spread in their
energy distributions, and therefore, may experience
the collective flavor transformation only when $n_\nu$ is large 
\cite{Duan:2005cp}.

\subsection{Slowly varying neutrino number density%
\label{sec:adiabatic-pendulum}}

If the neutrino number density $n_\nu$ varies with time,
Eq.~\eqref{eq:nfis-eom} is still valid but Eq.~\eqref{eq:SpSm-eom} 
is not. In this case,
Eq.~\eqref{eq:p-vartheta-eom} is
also valid as long as $n_\nu/n_\nu^0\gg1$.
We note that $\vartheta$ and $p_\vartheta$
comprise a canonically conjugate coordinate
and momentum pair. In these variables
the flavor pendulum has Hamiltonian
\begin{equation}
\mathcal{H}=\muV{}\left[\frac{1}{2}\left(\nratio\right)p_\vartheta^2
-\cos\vartheta\right].
\label{eq:Hamiltonian}
\end{equation}

Ref.~\cite{Hannestad:2006nj} first noticed that
the amplitude of  flavor mixing in this bipolar system (or
equivalently, the maximal angular position $\vartheta_{\max}$
of the flavor pendulum) decreases with the neutrino
number density $n_\nu$. Drawing an analogy to
the relation between the kinetic energy and
the angular momentum of a pirouette performer,
Ref.~\cite{Hannestad:2006nj}
suggested an intuitive explanation for this phenomenon:
As $n_\nu$ becomes smaller, the effective mass 
$m_\eff\equiv \muV{}^{-1}(n_\nu^0/n_\nu)$ in Eq.~\eqref{eq:Hamiltonian}
increases. As a result, the kinetic
energy of the flavor pendulum $E_\mathrm{kin}=p_\vartheta^2/(2m_\eff)$ is
reduced with smaller $n_\nu$ and the flavor pendulum cannot
swing as high as before.

We can arrive at the same conclusion for the scenarios
$n_\nu /n_\nu^0\gg1$ using Eq.~\eqref{eq:p-vartheta-eom}.
Let us compare the evolution of two flavor
pendulums (a) and (b). We assume that the two pendulums
have the same values of
$(\vartheta, p_\vartheta, \dot\vartheta, \dot{p}_\vartheta, n_\nu)$
at instant $t=t_0$. We also assume that
pendulum (a) has constant $n_\nu$ and that (b) has $n_\nu$
decreasing with time. After an infinitesimal interval $\Delta t$,
both pendulums will have the same values of
$(\vartheta, p_\vartheta, \dot{p}_\vartheta)$ but pendulum (b)
has smaller  $(\dot\vartheta, n_\nu)$ than (a) does
 [see Eq.~\eqref{eq:p-vartheta-eom}]. This is equivalent
to saying that both pendulums have the same angular position
and potential well but pendulum (b) possesses less 
kinetic energy  $E_\mathrm{kin}=\dot\vartheta p_\vartheta/2$ than (a) does.
As a result, pendulum (b) will not swing as high as (a) even
if $n_\nu$ is constant for $t>t_0+\Delta t$.

We note that neither Eq.~\eqref{eq:p-vartheta-eom} in this paper
or Eq.~(7) in Ref.~\cite{Hannestad:2006nj} is equivalent
to the original e.o.m.~of the NFIS's [Eq.~\eqref{eq:nfis-eom}]
if $n_\nu$ is small and varies with time.
Therefore, this explanation fails for $n_\nu/n_\nu^0\lesssim1$.

To quantify the relation between the maximal angular
position $\vartheta_{\max}$ of the flavor pendulum
and the neutrino number density $n_\nu$, we note that
in the limit $n_\nu/ n_\nu^0\gg1$
\begin{equation}
\mathcal{A}\equiv\oint p_\vartheta\,\ud\vartheta
\label{eq:action}
\end{equation}
is an adiabatic invariant of the pendulum motion
(see, \textit{e.g.}, Ref.~\cite{Landau:1976aa}).
The integration in Eq.~\eqref{eq:action} is performed over
one  pendulum cycle with 
$-\vartheta_{\max}\leq\vartheta\leq\vartheta_{\max}$. 
The neutrino number density $n_\nu$ is taken to be constant
during this cycle.

If $n_\nu$ is constant, the Hamiltonian of the flavor pendulum 
is also a constant and is $-\muV{}\cos\vartheta_{\max}$.
Using Eq.~\eqref{eq:Hamiltonian} we obtain
\begin{equation}
|p_\vartheta| = \sqrt{\frac{2 n_\nu^0}{n_\nu}}
\sqrt{\cos\vartheta-\cos\vartheta_{\max}}.
\label{eq:p-val}
\end{equation}
Combining Eqs.~\eqref{eq:action} and \eqref{eq:p-val} we have
\begin{subequations}
\begin{align}
\mathcal{A} 
&= -4\sqrt{\frac{2 n_\nu^0}{n_\nu}}\int_{\vartheta_{\max}}^0 
\sqrt{\cos\vartheta-\cos\vartheta_{\max}}\,\ud \vartheta \\
&= 16 \sqrt{\frac{n_\nu^0}{n_\nu}} W(\vartheta_{\max}).
\label{eq:action2}
\end{align}
\end{subequations}
Function $W(\xi)$ in Eq.~\eqref{eq:action2} is defined as
\begin{subequations}
\begin{align}
W(\xi) &\equiv \frac{1}{2\sqrt{2}} 
\int_0^{\xi}\sqrt{\cos\zeta-\cos\xi}\,\ud\zeta\\
&=\bm{E}(\sin(\xi/2))-\cos^2(\xi/2)\bm{K}(\sin(\xi/2)),
\end{align}
\end{subequations}
where
\begin{equation}
\bm{E}(k) \equiv \int_0^{\pi/2} \sqrt{1-k^2\sin^2\zeta} \,\ud\zeta
\end{equation}
is the complete elliptic integral of the second kind
\cite{Gradshteyn:1994aa}.
If $n_\nu/ n_\nu^0\gg1$ and $n_\nu$ varies slowly
(adiabatic process), then $\vartheta_{\max}$ as a function of time
satisfies the following relation:
\begin{equation}
W(\vartheta_{\max})\simeq W(2\thetav) \sqrt{\frac{n_\nu(t)}{n_\nu(0)}}.
\label{eq:vartheta-nnu-rel}
\end{equation}

\begin{figure}
\begin{center}
\includegraphics*[width=\myfigwid, keepaspectratio]{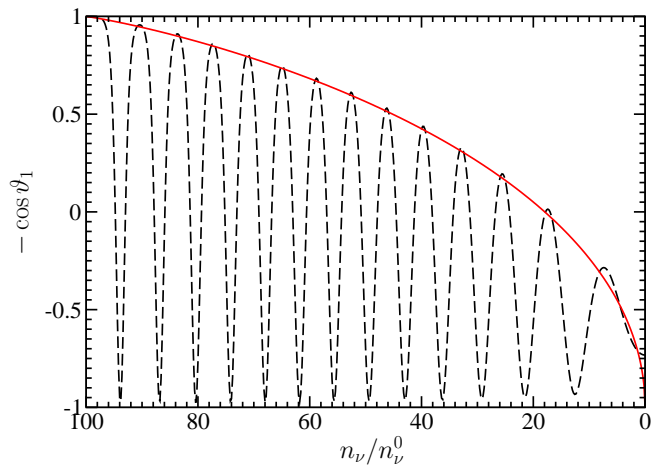}
\end{center}
\caption{\label{fig:pendulum-theta}(Color online)
Flavor oscillations of the simple symmetric bipolar system
in a nearly adiabatic process [Eq.~\eqref{eq:adiabatic-nnu} with
$\gamma=10$ and $n_\nu(0)/n_\nu^0=100$] for $\qthetav=0.01$. The dashed line
is the value of $-\cos\vartheta_1$ as a function $n_\nu$ computed from
Eq.~\eqref{eq:nfis-eom} with the initial conditions in Eq.~\eqref{eq:s12-ini}.
The solid line is $-\cos\vartheta_{\max}$ computed from 
Eq.~\eqref{eq:vartheta-nnu-rel}.}
\end{figure}

An interesting scenario is that $\tilde{\theta}_\mathrm{v}=\pi/2-\thetav\ll 1$. 
In this case, $\basef{z}\simeq-\HV$ and the probability
for a neutrino to be $\nu_e$ is
\begin{equation}
P_{\nu_e\nu_e} = \frac{1}{2}+\sB_1\cdot\basef{z}
\simeq \frac{1-\cos\vartheta_1}{2}
\lesssim \frac{1-\cos\vartheta_{\max}}{2},
\end{equation}
where $\vartheta_1$ is the angle between the directions
of $\sB_1$ and $\HV$.
Noting that $W(2\thetav)\simeq W(\pi)=1$, we find that 
\begin{equation}
0\lesssim P_{\nu_e\nu_e}(t)\leq (P_{\nu_e\nu_e})_{\max},
\end{equation}
where 
\begin{equation}
(P_{\nu_e\nu_e})_{\max}\simeq
\frac{1}{2}\left[1-\cos\left(W^{-1}
\left(\sqrt{\frac{n_\nu(t)}{n_\nu(0)}}\right)\right)\right]
\label{eq:Pnunu-max}
\end{equation}
is the maximal value that $P_{\nu_e\nu_e}$ may take
for a given $n_\nu$. In the above equation
 $W^{-1}(\xi)$ is the corresponding inverse function of $W(\xi)$.

For a concrete example, we assume that $n_\nu(t)$
has a linear dependence on time $t$:
\begin{equation}
n_\nu(t)=n_\nu(0)(1-\gamma^{-1} \muV{} t),
\label{eq:adiabatic-nnu}
\end{equation}
where $\gamma$ is the adiabatic parameter. The larger the value of
$\gamma$, the more adiabatic  the process is.
Taking $\qthetav=0.01$, we calculate the value of
$-\cos\vartheta_1$ as a function of $n_\nu$
for $\gamma=10$ and $n_\nu(0)/\nu_\nu^0=100$
by solving Eq.~\eqref{eq:nfis-eom} with the 
initial conditions in Eq.~\eqref{eq:s12-ini}.
The results are plotted as the dashed line in 
Fig.~\ref{fig:pendulum-theta}.
We have also computed $\vartheta_{\max}$ from Eq.~\eqref{eq:vartheta-nnu-rel}
and plot $-\cos\vartheta_{\max}$
as the solid line in Fig.~\ref{fig:pendulum-theta}.
It is clear that $-\cos\vartheta_{\max}$ outlines the upper envelope
of $-\cos\vartheta_1$ for $n_\nu/ n_\nu^0\gg1$.

Using the analogy of harmonic oscillators, Ref.~\cite{Hannestad:2006nj} 
has argued that, for the scenario with $\qthetav\ll1$, 
$(P_{\nu_e\nu_e})_{\max}$ should depend linearly on $\sqrt{n_\nu}$,
at least when  $\vartheta_{\max}\ll 1$. However,
it is clear that this conjecture is not true if $\vartheta_{\max}$ 
is significant. In this case,  $(P_{\nu_e\nu_e})_{\max}$
can be understood using the general form
of $W(\xi)$ and Eq.~\eqref{eq:Pnunu-max}.
On the other hand, we note that
\begin{equation}
W(\vartheta_{\max}) \simeq 
\frac{\pi}{4}\left(\frac{\vartheta_{\max}}{2}\right)^2
\label{eq:W-small-xi}
\end{equation}
for $\vartheta_{\max}\ll 1$ \cite{Gradshteyn:1994aa}.
Using Eqs.~\eqref{eq:vartheta-nnu-rel} and \eqref{eq:W-small-xi}
we obtain
\begin{equation}
\frac{\vartheta_{\max}^2(t)}{\vartheta_{\max}^2(t_0)}
\simeq \sqrt{\frac{n_\nu(t)}{n_\nu(t_0)}},
\end{equation}
where $t_0$ is an instant at which $\vartheta_{\max}(t_0)\ll1$. Because 
$P_{\nu_e\nu_e}\simeq (1-\cos\vartheta_{1})/2\simeq \vartheta_{1}^2/4$,
we have
\begin{equation}
(P_{\nu_e\nu_e})_\mathrm{\max} \simeq \frac{\vartheta_{\max}^2(t_0)}{4}
\sqrt{\frac{n_\nu}{n_\nu(t_0)}} 
\end{equation}
for $P_{\nu_e\nu_e}\ll1$.
Therefore, $(P_{\nu_e\nu_e})_{\max}$  does depend linearly on 
$\sqrt{n_\nu}$  in the limit   $\vartheta_{\max}\ll 1$. 
Note that this result only applies for $n_\nu/n_\nu^0\gg1$.
As mentioned above, our argument about the adiabatic invariant
fails for $n_\nu/n_\nu^0\lesssim1$.

\section{Asymmetric Bipolar System%
\label{sec:asymmetric-sys}}

\subsection{Gyroscopic flavor pendulum\label{sec:top-eom}}

We now consider a simple asymmetric bipolar system
initially consisting of mono-energetic $\nu_e$ and $\bar\nu_e$
with different but constant 
number densities. We note that Eq.~\eqref{eq:nfis-eom}
is still valid
except that we now take $n_{\nu,1}=n_{\nu}$ and $n_{\nu,2}=\alpha n_\nu$
with $\alpha\neq1$ being a positive constant.
Ref.~\cite{Hannestad:2006nj} has shown that
this asymmetric bipolar system is equivalent to a 
gyroscopic pendulum or a spinning top in flavor space for which
\begin{subequations}
\label{eq:S12-ini}
\begin{align}
\Sb_+(0) &= \frac{1-\alpha}{2}n_\nu\basef{z}\\
\Sb_-(0)&=\frac{1+\alpha}{2}n_\nu\basef{z}.
\end{align}
\end{subequations}
To see this we define
\begin{equation}
\mathbf{Q} \equiv \Sb_- - \frac{\muV{}}{\mu_\nu}\HV.
\label{eq:Q-def}
\end{equation}
(Although we follow Ref.~\cite{Hannestad:2006nj} in demonstrating
the equivalence of an asymmetric bipolar system and a gyroscopic pendulum,
we have adopted somewhat different notations for our convenience.)
Using Eqs.~\eqref{eq:Sm-eom} and \eqref{eq:Q-def} one sees that 
 $\mathbf{Q}$ obeys the e.o.m.
\begin{equation}
\frac{\ud}{\ud t}\Qb = \mu_\nu {\Qb} \times \Sb_+
\label{eq:Q-eom}
\end{equation}
and maintains a constant magnitude
\begin{equation}
\begin{split}
Q = n_\nu \left(\frac{1+\alpha}{2}\right)  
\Bigg[&1
+\left(\frac{4}{1+\alpha}\right)\left(\frac{n_\nu^0}{n_\nu}\right)
\cos2\thetav\\
&+\left(\frac{2}{1+\alpha}\right)^2\left(\frac{n_\nu^0}{n_\nu}\right)^2
\Bigg]^{1/2}.
\end{split}
\end{equation}

With the definition of
\begin{subequations}
\begin{align}
\Jb &\equiv \frac{\Sb_+}{n_\nu},\\
\rB &\equiv \frac{\Qb}{Q},
\end{align}
\end{subequations}
Eqs.~\eqref{eq:Sp-eom}, \eqref{eq:Q-def} and \eqref{eq:Q-eom} lead to
\begin{subequations}
\label{eq:J-r-eom}
\begin{align}
\dot{\Jb} &= \frac{\muV{}Q}{n_\nu}\rB \times \HV,
\label{eq:J-eom}\\
\dot{\rB} &= -\muV{}\left(\nratio\right) \rB\times\Jb.
\label{eq:r-eom}
\end{align}
\end{subequations}
Using Eq.~\eqref{eq:J-r-eom} one can easily show that 
\begin{equation}
\sigma\equiv\Jb\cdot\rB
\end{equation}
is a constant of motion.
From Eq.~\eqref{eq:r-eom} one obtains
\begin{equation}
\rB\times\dot{\rB}=-\muV{}\left(\nratio\right) (\sigma\rB-\Jb).
\label{eq:r-rdot}
\end{equation}
We note that Eqs.~\eqref{eq:J-eom} and \eqref{eq:r-rdot} are
equivalent to
\begin{subequations}
\label{eq:spinning-top}
\begin{align}
\dot{\Jb} &= \rB \times M \gB,\\
\Jb &= M \rB \times \dot{\rB} +\sigma\rB,
\label{eq:J-val}
\end{align}
\end{subequations}
where
\begin{align}
M &= \frac{1}{\muV{}}\left(\frac{n_\nu^0}{n_\nu}\right),
\label{eq:M}\\
\gB &= \frac{\muV{}^2}{n_\nu^0}Q\HV.
\end{align}
Therefore, this asymmetric bipolar system is indeed equivalent
to a gyroscopic flavor pendulum.
Specifically, $\rB$ is the position vector of the bob,
 $\Jb$ is the total angular momentum, $\sigma$ is the internal
angular momentum of the bob, $M$ is the mass of the bob,
and $\gB$ is the acceleration field.
The only difference
between this pendulum and that shown in Fig.~\ref{fig:pendulum}(a)
is the spin of the bob. Hereafter we will
loosely refer to both the symmetric and asymmetric
bipolar systems as flavor pendulums.

The motion of the gyroscopic flavor pendulum is the combination
of a precession around $\HV$  and 
a nutation with $\vartheta_{\min}\leq\vartheta\leq\vartheta_{\max}$.
Here $\vartheta$ is  the polar angle of $\rB$ with respect to
$\basev{z}=\HV$, and $\vartheta_{\min}$ and $\vartheta_{\max}$
are the minimal and maximal values of $\vartheta$ during nutation.
For the simple asymmetric bipolar system that we have
discussed, one has $\vartheta_{\max}=\vartheta|_{t=0}$. 
The value of $\vartheta_{\min}$
can be determined as follows.
Following Ref.~\cite{Hannestad:2006nj} we define the
total energy of the pendulum as
\begin{subequations}
\label{eq:top-energy}
\begin{align}
E_\mathrm{tot}&= E_\mathrm{pot} + E_\mathrm{kin}\\
&=Mg(1-\HV\cdot\rB) + \frac{\Jb^2}{2M}\\
&=Mg(1-\cos\vartheta) + \frac{M\dot{\rB}^2}{2} + \frac{\sigma^2}{2M}.
\end{align}
\end{subequations}
We note that $E_\mathrm{tot}$ differs from
the conserved total effective energy of the NFIS's
\cite{Duan:2005cp} by only a
constant multiplicative factor and a additive constant,
and therefore, is also conserved.
Because the motion of the pendulum
is a pure precession around $\HV$ when $\vartheta=\vartheta_{\min}$,
one has
\begin{equation}
\label{eq:E-conservation}
Mg(\cos\vartheta_{\min}-\cos\vartheta_{\max})
=\frac{1}{2}M\sin^2\vartheta_{\min}\dot{\varphi}^2|_{\vartheta=\vartheta_{\min}},
\end{equation}
where $\varphi$ is the azimuthal angle of $\rB$ with 
respect to $\HV$. Using the conservation
of the total angular momentum 
in the direction of the acceleration field $\gB$, one obtains
\begin{equation} 
\label{eq:Jz-conservation}
\sigma(\cos\vartheta_{\min}-\cos\vartheta_{\max})
=-M\sin^2\vartheta_{\min}\,\dot{\varphi}|_{\vartheta=\vartheta_{\min}}.
\end{equation}
Combining Eqs.~\eqref{eq:E-conservation} and \eqref{eq:Jz-conservation} 
we have
\begin{equation}
\cos\vartheta_{\min}
=-\eta+\sqrt{(\eta+\cos\vartheta_{\max})^2+(1-\cos^2\vartheta_{\max})},
\label{eq:vartheta-min}
\end{equation}
where
\begin{equation}
\eta \equiv \frac{\sigma^2}{4M^2g}.
\label{eq:eta}
\end{equation}

\subsection{Precession/nutation modes and synchronized/bipolar regimes%
\label{sec:two-modes}}

We shall refer to neutrino flavor transformation
as being in the precession (nutation) mode when the 
corresponding analogous flavor pendulum
is undergoing precession (nutation).
The symmetric bipolar system discussed in Sec.~\ref{sec:symmetric-sys}
corresponds to the limit $\alpha=1$ and is always in a pure
nutation mode. In this limit $\sigma=0$ and $\vartheta_{\min}=0$, 
so the pendulum does not spin at all and simply swings in a fixed plane.
Taking $p_\vartheta=\Jb\cdot\basev{y}$ and $\vartheta$ as the angle
between $\rB$ and $\HV$, one can obtain from Eq.~\eqref{eq:J-r-eom} that
\cite{Hannestad:2006nj}
\begin{subequations}
\label{eq:p-vartheta-exact}
\begin{align}
\dot{p}_\vartheta &= -\muV{}\left(\frac{Q}{n_\nu}\right)\sin\vartheta\\
\dot{\vartheta} &= \muV{}\left(\frac{n_\nu}{n_\nu^0}\right) p_\vartheta.
\end{align}
\end{subequations}
Eq.~\eqref{eq:p-vartheta-exact} is the exact version of
Eq.~\eqref{eq:p-vartheta-eom}. In the limit
$n_\nu/n_\nu^0\gg1$, $\Qb\simeq\Sb_-$ and
$\Sb_-$ approximately follows a plane pendulum motion as we have
discussed in Sec.~\ref{sec:pendulum-eom}.
If the neutrino number density $n_\nu$ is constant and
the vacuum mixing angle $\thetav$ or $\qthetav$ is small,
the bipolar systems in the pure nutation mode 
can experience almost complete flavor conversion during a nutation period.
This is true for various initial configurations 
(see Table I in Ref.~\cite{Duan:2005cp}).

A bipolar system generally evolves simultaneously in both
precession and nutation modes. However, if 
the neutrino number density is large enough,
it has been shown that a neutrino gas is 
in the synchronized mode with a characteristic frequency $\wsync$ 
independent of its initial configuration \cite{Pastor:2001iu}.
The criterion for synchronization can be written as \cite{Duan:2005cp}
\begin{equation}
|\mu_\nu \Sb|\gg|\langle\muV{}\rangle|,
\label{eq:syn-general}
\end{equation}
where
\begin{equation}
\Sb\equiv \sum_i n_{\nu,i}\sB_i
\label{eq:NFIS-tot}
\end{equation}
is the total NFIS, and
\begin{equation}
\langle\muV{}\rangle\equiv
\sum_i\frac{\muV{,i}n_{\nu,i}\sB_i\cdot\Sb}{\Sb^2}=\wsync
\label{eq:avg-muV}
\end{equation}
is the average vacuum oscillation frequency.
The index $i$ in Eqs.~\eqref{eq:NFIS-tot} and \eqref{eq:avg-muV}
denotes neutrinos or antineutrinos with a specific momentum.

For the initial condition in Eq.~\eqref{eq:S12-ini} we note that,
if $n_\nu$ is large, the total
angular momentum of the flavor pendulum is dominated by its spin 
\begin{equation}
\Jb\simeq\sigma\rB
\label{eq:J-large-nnu}
\end{equation}
and 
\begin{equation}
\sigma\simeq\frac{1-\alpha}{2}.
\label{eq:sigma-large-nnu}
\end{equation}
In the limit
\footnote{Ref.~\cite{Hannestad:2006nj} first noticed that the 
flavor pendulum
with $\qthetav\ll1$ possesses little nutation for $n_\nu\gg\nnux$ and
is in the synchronized regime. 
We note that a flavor pendulum will have little nutation
so long as Eq.~\eqref{eq:syn-limit} is satisfied.
This result is independent of the value of $\thetav$.}
\begin{equation}
n_\nu\gg\nnux\equiv 8 \frac{1+\alpha}{(1-\alpha)^2} n_\nu^0,
\label{eq:syn-limit}
\end{equation}
the parameter $\eta$ [see Eq.~\eqref{eq:eta}]
satisfies $\eta\gg 1$ and 
\begin{equation}
\cos\vartheta_{\min}
\simeq-\eta+\eta\left(1+\frac{\cos\vartheta_{\max}}{\eta}\right)
\simeq\cos\vartheta_{\max}.
\end{equation}
As a result, the flavor pendulum roughly maintains a constant latitude
and is essentially in the precession mode.
One can explicitly show that in this case the flavor
pendulum precesses around $-\HV$ with a constant
angular frequency \cite{Hannestad:2006nj}
\begin{equation}
\label{eq:wsync}
\Omega\simeq\wsync\equiv\frac{1+\alpha}{1-\alpha}\muV{}.
\quad
\end{equation}
Therefore,
a bipolar system is synchronized if $n_\nu\gg\nnux$ and the
synchronized mode corresponds to a pure precession mode with
the synchronization frequency $\wsync$ as its precession frequency.
We refer to the limit in Eq.~\eqref{eq:syn-limit} as the
``synchronized regime''. We say that a bipolar system is 
in the ``bipolar regime'' if Eq.~\eqref{eq:syn-limit}
is not satisfied. In this case it can be in both the precession
and nutation modes.

For the simple asymmetric bipolar system discussed here, 
Eq.~\eqref{eq:syn-general} lead to
\begin{equation}
n_\nu\gg 2 \frac{1+\alpha}{(1-\alpha)^2}n_\nu^0 = \frac{\nnux}{4}.
\label{eq:syn-limit2}
\end{equation}
Eqs.~\eqref{eq:syn-limit} and \eqref{eq:syn-limit2} differ 
by a constant multiplicative factor. This reflects
the fact that there is \textit{no} sharp boundary
between the synchronized and bipolar regimes.
The simple prescription for
the synchronization frequency $\wsync$ in Eq.~\eqref{eq:avg-muV} 
allows a ready and practical application of 
the synchronization condition in Eq.~\eqref{eq:syn-general} 
 for neutrino and/or antineutrino gases with finite
spreads in their energy spectra.

The ways in which the word ``bipolar'' has been used in the literature
\cite{Duan:2005cp,Duan:2006an,Duan:2006jv,Hannestad:2006nj,Raffelt:2007yz}
can be very confusing. There is a tendency to mistakenly identify
the synchronized (bipolar) regime with the precession (nutation) mode.
This is probably because a flavor pendulum can only precess in
the synchronized regime and a symmetric bipolar system
was once viewed as a typical bipolar system  which
is always in a nutation mode.
However, the criterion determining whether a bipolar system
is mostly in the precession or nutation mode is not the same
as that for determining whether it is in the synchronized or bipolar regime.
A good  example is that an asymmetric bipolar system
can simultaneously be in both the precession and
nutation modes in the bipolar regime. We also note that, 
while the precession frequency $\Omega$ of the 
precession mode in the synchronized regime 
is determined from Eq.~\eqref{eq:avg-muV}
and is independent of the neutrino number density $n_\nu$,
the precession frequency of a precession mode in the bipolar regime
depends on  $n_\nu$ (see Sec.~\ref{sec:omega}).

The evolution of a bipolar system in the bipolar regime is
also referred to as bimodal oscillations.
Ref.~\cite{Samuel:1995ri} has shown that,
in an asymmetric bipolar system initially consisting
of mono-energetic $\nu_e$ and $\bar\nu_e$,
the $x$- and $y$-components of the polarization
vectors of the neutrino and antineutrino ($2\sB_1$ and $-2\sB_2$ in
the NFIS notation) are bimodal as they are functions
of two intrinsic periods.
It is clear that these two periods 
are related to the precession and nutation of the flavor pendulum.
If $\nu_e$ and $\bar\nu_e$ have different energies or the system
starts with different neutrino/antineutrino species, one can demonstrate
that such a system is equivalent to a flavor pendulum
in some properly chosen corotating frame \cite{Duan:2005cp}.
In this case the precession frequency $\Omega$ is shifted 
by the rotation frequency of the corotating frame. 

\section{Pure Precession Mode of Asymmetric Bipolar Systems%
\label{sec:precession-mode}}

\subsection{Pure precession mode%
\label{sec:pure-precession}}

\begin{figure*}
\begin{center}
\begin{center}
$\begin{array}{@{}c@{\hspace{\myfigsep}}c@{}}
\includegraphics*[width=\myfigwid, keepaspectratio]{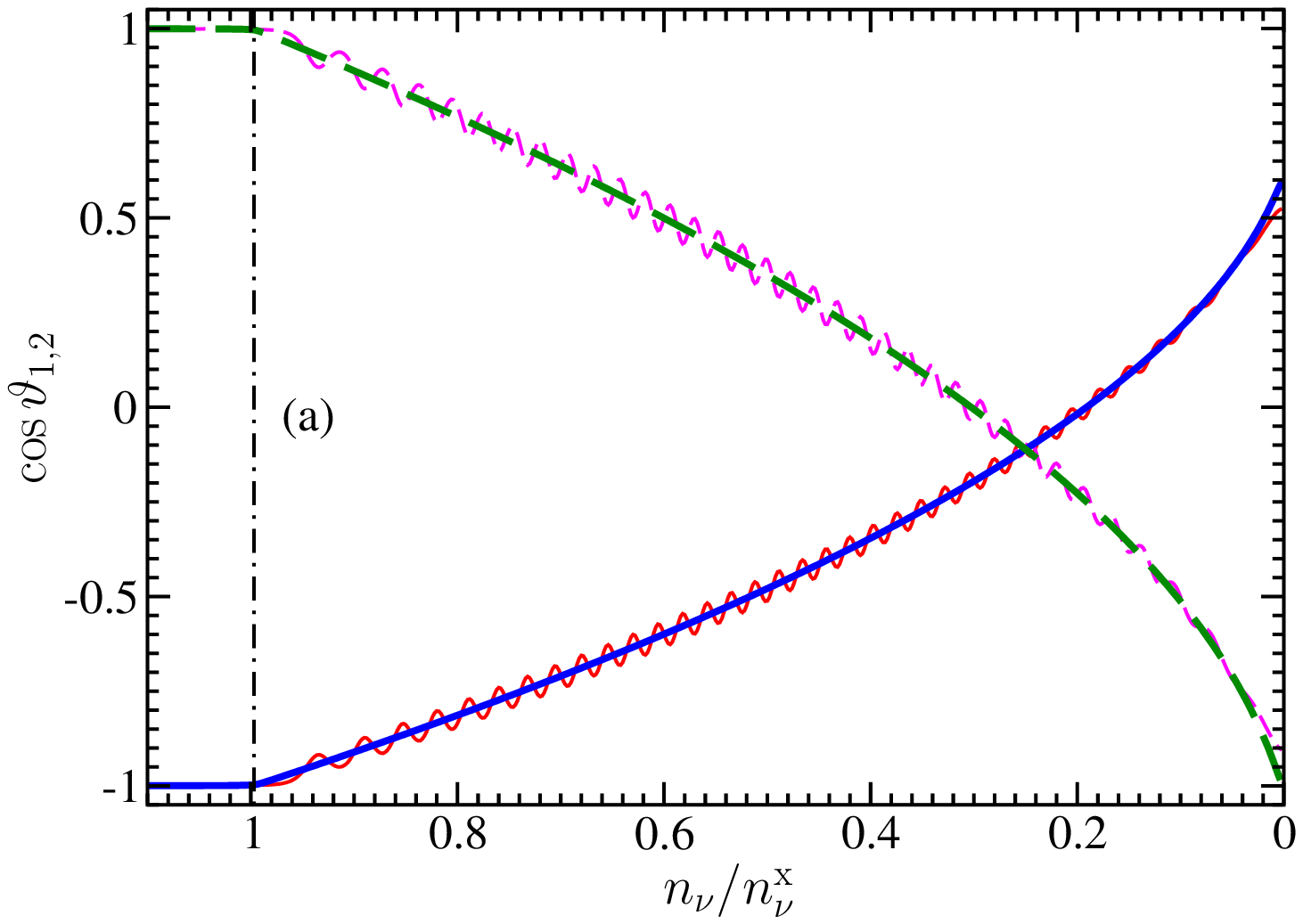} &
\includegraphics*[width=\myfigwid, keepaspectratio]{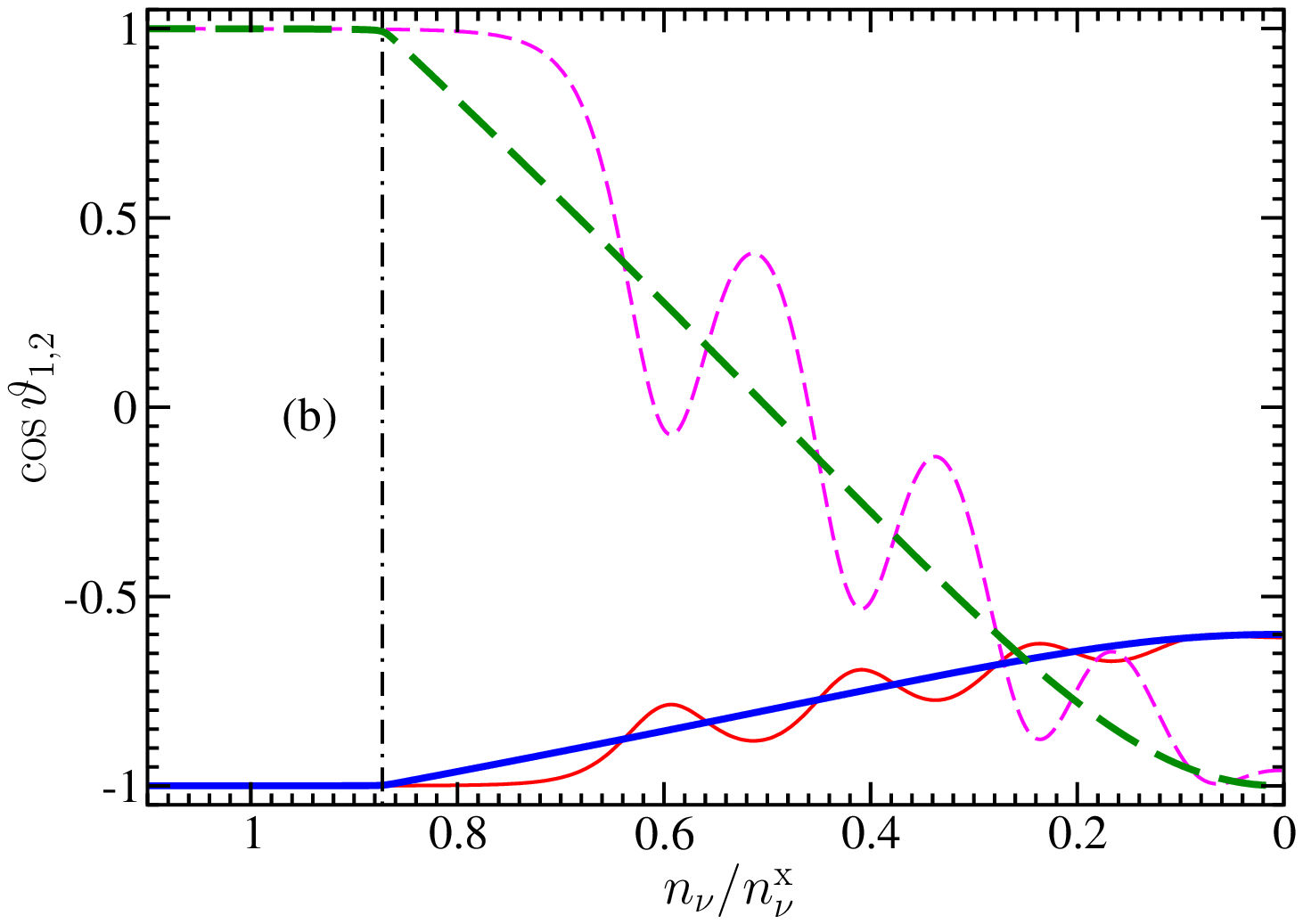} \\
\includegraphics*[width=\myfigwid, keepaspectratio]{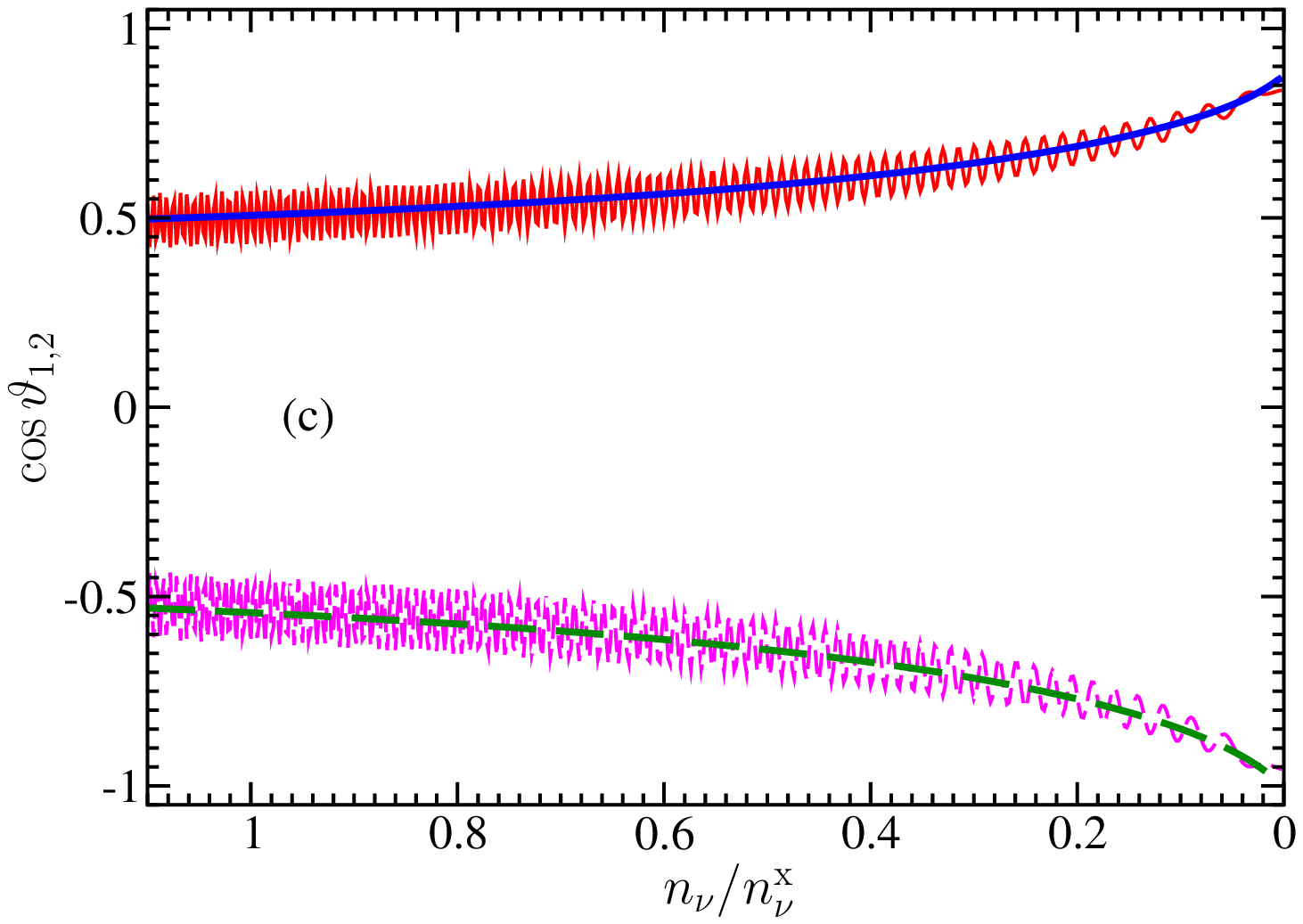} &
\includegraphics*[width=\myfigwid, keepaspectratio]{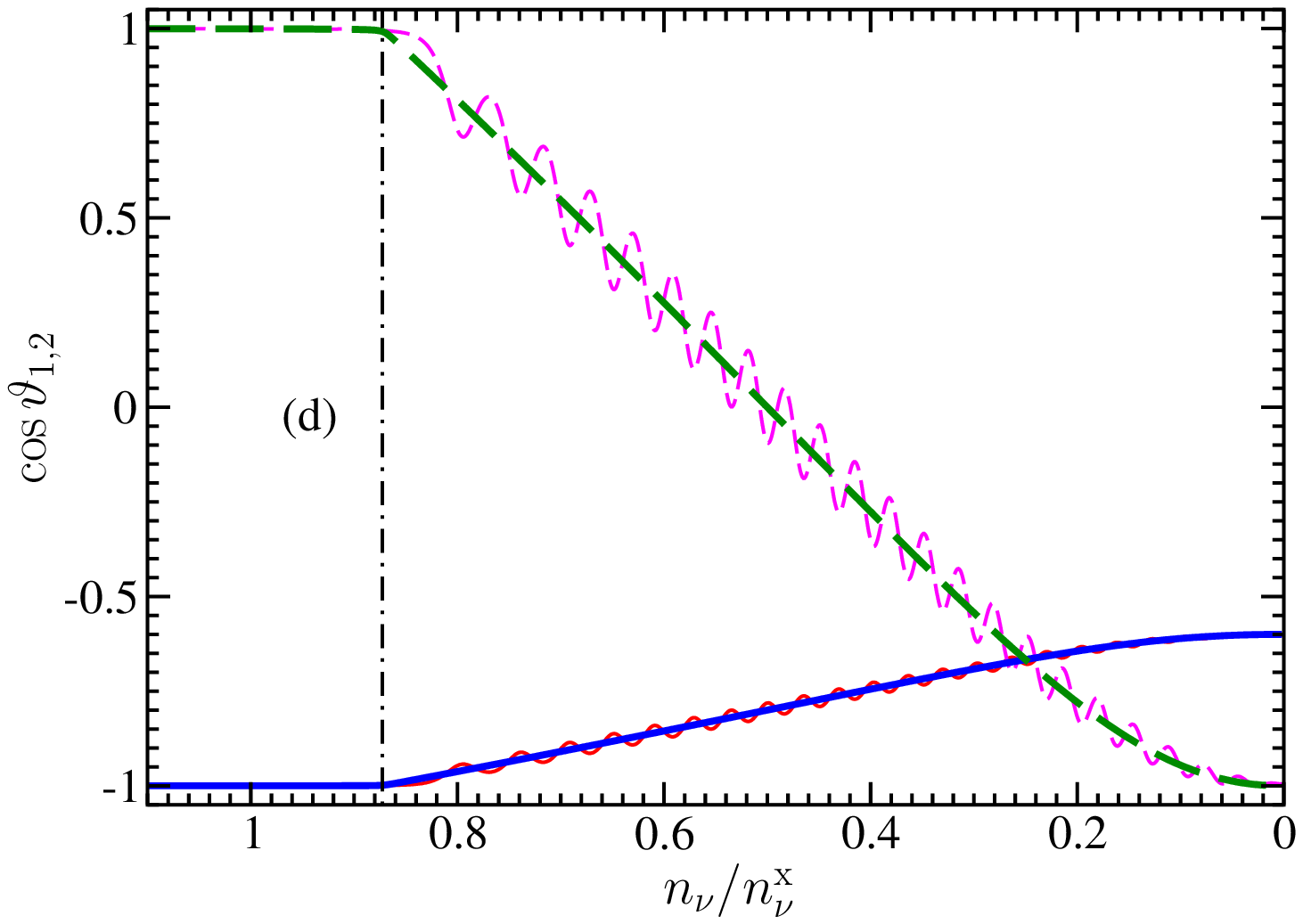}
\end{array}$
\end{center}
\end{center}
\caption{\label{fig:top-theta}(Color online)
Values of $\cos\vartheta_{1}$ (solid lines) 
and $\cos\vartheta_{2}$ (dashed lines),
where $\vartheta_1$ and $\vartheta_2$ are
the polar angles of the NFIS's $\sB_1$ and $\sB_2$ with respect to $\HV$, 
as functions of neutrino number density $n_\nu$
for simple asymmetric bipolar systems.
Panels (a), (b) and (d) have mixing angle $\qthetav=0.01$ and
panel (c) has $\thetav=0.6$. The asymmetry parameter is $\alpha=0.8$
for (a) and (c), and $\alpha=0.2$ for (b) and (d).
The thin lines are computed from the original e.o.m.~of 
NFIS's $\sB_1$ and $\sB_2$ [Eq.~\eqref{eq:nfis-eom}
with the initial conditions in \eqref{eq:s12-ini}] assuming a nearly adiabatic
process [Eq.~\eqref{eq:adiabatic-nnu} with
 $n_\nu(0)/\nnux=2$]. The adiabatic parameter is 
$\gamma=40$ for panels (a--c) and
is $200$ for panel (d). The thick lines are
computed from Eq.~\eqref{eq:vartheta-12} assuming
that bipolar systems stay in the pure precession mode.
The vertical dot-dashed lines in panels (a), (b) and (d)
correspond to $n_\nu=\nnuc$.}
\end{figure*}

Although a bipolar system tends to develop
some nutation  in addition to the precession mode
in the bipolar regime,
the actual mix of these modes depends on
the initial conditions as well as the
system configuration. For example,
a flavor pendulum precesses around
$-\HV$ with constant angular frequency $\Omega$
without any nutation if
\begin{subequations}
\label{eq:precession-cond}
\begin{align}
Mg\sin\vartheta
&= \Omega J_\perp \\
&=\Omega(M\Omega\sin\vartheta\cos\vartheta+\sigma\sin\vartheta)
\end{align}
\end{subequations} 
is satisfied, and the corresponding bipolar system
is in the pure precession mode. For a gyroscope this
is known as the ``regular precession''.

With varying neutrino number densities the problem is generally
complicated. This is because 
almost all the parameters of the flavor
pendulum ($Q$, $M$, $g$, $\sigma$, \textit{etc.}) depend on $n_\nu$
and the e.o.m.~of a pendulum, Eq.~\eqref{eq:spinning-top},
is not equivalent to that of the NFIS's 
if $n_\nu(t)$ is not constant.
In this case, one has to use Eq.~\eqref{eq:nfis-eom}
to follow the evolution of the bipolar system.
Simple numerical examples presented in Ref.~\cite{Hannestad:2006nj} 
seem to suggest that the evolution
of a bipolar system with $\thetav\simeq\pi/2$ 
can be dominantly in a precession mode
after the system transitions from the synchronized regime into the 
bipolar regime. Here we try to gain some analytical understanding
of this precession mode by using the same simple bipolar system
studied in Sec.~\ref{sec:asymmetric-sys} but with time-varying
$n_\nu$.

We note that, in the synchronized regime (\textit{i.e.}, the limit
of large $n_\nu$), both $\sB_1$ and
$\sB_2$ precess uniformly around $\HV$, and the motion of
the NFIS's has a cylindrical symmetry around the axis
along $\HV$. This symmetry is inherited from
the e.o.m.~of the NFIS's [Eq.~\eqref{eq:nfis-eom}].
We consider an infinitely long process during which
$n_\nu$ is decreased without preference to any azimuthal
angle with respect to $\HV$. The cylindrical symmetry
in the motion of the NFIS's around $\HV$ is expected to be 
preserved in such a process, and $\sB_1$ and $\sB_2$
keep on precessing uniformly around $\HV$ without any wobbling.

If this is true,
vectors $\sB_1$, $\sB_2$ and $\HV$ must always be in the same plane,
and $\sB_1$ and $\sB_2$ rotates around $-\HV$ with the same angular frequency
\begin{subequations}
\label{eq:omega}
\begin{align}
\Omega 
&= \muV{}\left[1-\frac{\alpha}{2\sin\vartheta_1}\left(\nratio\right)
\sin(\vartheta_1+\vartheta_2)\right]\\
&= \muV{}\left[-1-\frac{1}{2\sin\vartheta_2}\left(\nratio\right)
\sin(\vartheta_1+\vartheta_2)\right],
\label{eq:omega-2}
\end{align}
\end{subequations}
where $\vartheta_{1(2)}$ is the angle between $\sB_{1(2)}$
and $\HV$.
On the other hand, from Eq.~\eqref{eq:nfis-eom}
it can be shown that $\HV\cdot(\sB_1+\alpha\sB_2)$
is time invariant even if $n_\nu$ changes with time. 
Consequently, we obtain the following
two equations for $\vartheta_1$ and $\vartheta_2$:
\begin{subequations}
\label{eq:vartheta-12}
\begin{align}
4\sin\vartheta_1\sin\vartheta_2
&=-\frac{n_\nu}{n_\nu^0}(\sin\vartheta_1-\alpha\sin\vartheta_2)
\nonumber\\
&\quad\times\sin(\vartheta_1+\vartheta_2),
\label{eq:s1-s2}\\
(1-\alpha)\cos2\thetav
&=\cos\vartheta_1+\alpha\cos\vartheta_2.
\label{eq:c1-c2}
\end{align}
\end{subequations}

We have solved Eq.~\eqref{eq:vartheta-12} numerically for  simple
asymmetric bipolar systems with different choices of $\thetav$
and asymmetry parameter $\alpha$. 
The results are plotted in Fig.~\ref{fig:top-theta}. 
For comparison, we have
also solved numerically 
the original e.o.m.~of the NFIS's, Eq.~\eqref{eq:nfis-eom}, 
 for the same 
bipolar systems assuming that $n_\nu$ changes in the way described
by Eq.~\eqref{eq:adiabatic-nnu}. These results are also shown in
Fig.~\ref{fig:top-theta}. Clearly, the polar angles $\vartheta_1$
and $\vartheta_2$ of the NFIS's $\sB_1$ and $\sB_2$
oscillate around those values determined from 
Eq.~\eqref{eq:vartheta-12} as $n_\nu$ decreases.
This is    true not only for the bipolar systems with $\thetav\simeq\pi/2$ but
also for those with other vacuum mixing angles. 

The results shown in Fig.~\ref{fig:top-theta} 
can be understood as follows.
Although Eqs.~\eqref{eq:nfis-eom} and \eqref{eq:spinning-top}
are not equivalent over a long period
for a time-varying $n_\nu$, we may
still view a bipolar system as a flavor pendulum
over a short time interval during which $n_\nu$ does not change much.
Suppose that at instant $t_1$ the
flavor pendulum precesses uniformly around $\HV$ at
 latitude $\vartheta_0(t_1)$. 
In the adiabatic limit this precession continues as $n_\nu$
slowly changes, but the value of $\vartheta_0$
changes with $n_\nu$ [$\vartheta_0$ is a function
of  $\vartheta_1$ and $\vartheta_2$ which vary
with $n_\nu$ according to Eq.~\eqref{eq:vartheta-12}].
Of course, in realistic conditions, $n_\nu$ can
only decrease with a finite rate, and
the actual polar angle $\vartheta$ of the pendulum always
``wobbles'' (as a result of excitation of nutation modes)
around $\vartheta_0$ with some nutation period $T_\mathrm{nut}$.
However, if $n_\nu$ changes so slowly that
\begin{equation}
\left[T_\mathrm{nut}\left(\frac{\ud n_\nu}{\ud t}\right)\right]^{-1}\gg
\left(\frac{\ud \vartheta_0}{\ud n_\nu}\right),\,
\left(\frac{\partial \vartheta}{\partial n_\nu}\right),
\label{eq:pure-precession-cond}
\end{equation}
$\vartheta$ can be expected to closely follow $\vartheta_0$,
and Eq.~\eqref{eq:vartheta-12} becomes an excellent
approximation. This expectation can be verified by
comparing panels (b) and (d) of Fig.~\ref{fig:top-theta}
where the evolution of two otherwise identical bipolar systems is
calculated using different adiabatic parameters 
[$\gamma=40$ in (b) and $200$ in (d)].
With a much slower change in $n_\nu$ for panel (d), the
result obtained by solving Eq.~\eqref{eq:vartheta-12}
becomes very close to that derived from the exact numerical
calculations.

The final values of $\vartheta_1|_{n_\nu=0}$ and $\vartheta_2|_{n_\nu=0}$ 
of the bipolar system
in the pure precession mode can be obtained as follows. 
The precession
frequency $\Omega$ of the flavor pendulum cannot be 0,
and therefore, cannot change sign as $n_\nu$ decreases.
Because $\Omega=\wsync>0$ 
in the limit of large $n_\nu$ for  $\alpha<1$ 
[Eq.~\eqref{eq:wsync}], we have $\Omega>0$.
Eq.~\eqref{eq:omega-2} would give $\Omega>0$ for $n_\nu=0$
only if
\begin{equation}
\vartheta_2|_{n_\nu=0}=\pi.
\label{eq:vartheta2-n0}
\end{equation}
This and Eq.~\eqref{eq:omega} then give
\begin{equation}
\Omega|_{n_\nu=0}=\muV{}.
\label{eq:omega-n0}
\end{equation}
Combining Eqs.~\eqref{eq:vartheta2-n0} and \eqref{eq:c1-c2} we obtain
\begin{equation}
\cos\vartheta_1|_{n_\nu=0}=(1-\alpha)\cos2\thetav+\alpha.
\end{equation}
This agrees with the numerical results shown in Fig.~\ref{fig:top-theta}.
For the inverted mass hierarchy scenario with $\thetav\simeq\pi/2$,
we have $\basef{z}\simeq-\HV$ and
\begin{subequations}
\begin{align}
P_{\nu_e\nu_e}|_{n_\nu=0}
&\simeq\left.\frac{1-\cos\vartheta_1}{2}\right|_{n_\nu=0}
\simeq 1-\alpha,\\
P_{\bar\nu_e\bar\nu_e}|_{n_\nu=0}
&\simeq\left.\frac{1+\cos\vartheta_2}{2}\right|_{n_\nu=0}
\simeq 0.
\end{align}
\end{subequations}
However, we note that these results for $n_\nu=0$ do not apply to
realistic bipolar systems with finite spreads in the neutrino
and antineutrino energy spectra as collective oscillations
of these systems always break down before $n_\nu$ reaches 0.

\subsection{Critical neutrino number density for the 
inverted mass hierarchy scenario with $\thetav\simeq\pi/2$%
\label{sec:nnuc}}

In Fig.~\ref{fig:top-theta} one can see that, for the inverted
mass hierarchy scenario with $\thetav\simeq\pi/2$, $\sB_1$
and $\sB_2$ begin to  misalign with $\HV$ when
$n_\nu$ is smaller than some critical value $\nnuc$,
and there seems to be discontinuity in $\ud \vartheta_{1(2)}/\ud n_\nu$
at $n_\nu=\nnuc$. To understand these results, we consider the limit where
$\thetav=\pi/2$.
We define
\begin{equation}
\label{eq:x1-x2-def}
x_{1(2)}\equiv\sin\vartheta_{1(2)}.
\end{equation} 
For $\vartheta_1\simeq\pi$ and $\vartheta_2\simeq0$ we have
\begin{subequations}
\label{eq:c12-small}
\begin{align}
\cos\vartheta_1
&\simeq-1+\frac{x_1^2}{2},\\
\cos\vartheta_2
&\simeq 1-\frac{x_2^2}{2}.
\end{align}
\end{subequations}
Combining Eqs.~\eqref{eq:vartheta-12} and \eqref{eq:c12-small}
we obtain
\begin{subequations}
\label{eq:x1-x2}
\begin{align}
4x_1 x_2 &\simeq (x_1 -\alpha x_2)\frac{n_\nu}{n_\nu^0} 
\nonumber \\
&\quad\times
\left[x_2\left(1-\frac{x_1^2}{2}\right)
-x_1\left(1-\frac{x_2^2}{2}\right)\right],\\
 1-\alpha &\simeq
\left(1-\frac{x_1^2}{2}\right)-\alpha\left(1-\frac{x_2^2}{2}\right).
\end{align}
\end{subequations}
Eq.~\eqref{eq:x1-x2} has the solution
\begin{equation}
x_1^2\simeq \alpha x_2^2
\simeq 2\sqrt{\alpha} \left(\frac{\nnuc}{n_\nu}-1\right)
\quad \mathrm{if}\,n_\nu<\nnuc,
\label{eq:x1-x2-sol}
\end{equation}
where 
\begin{equation}
\nnuc \equiv \frac{4 n_\nu^0}{(1-\sqrt{\alpha})^2}.
\label{eq:nnuc}
\end{equation}
Therefore, for the limiting case with $\thetav=\pi/2$,
$\sB_1$ and $\sB_2$ only start to misalign with $\HV$
when $n_\nu$ becomes smaller than $\nnuc$.

One can also obtain the same value for $\nnuc$
from the gyroscopic flavor pendulum analogy using
Eq.~\eqref{eq:vartheta-min}. For $\vartheta_{\max}=\pi$,
one always has $\vartheta_{\min}=\pi$ if
\footnote{\label{fn:nnuc}Ref.~\cite{Hannestad:2006nj}
pointed out the existence of the critical
neutrino number density $\nnuc$ using this argument
but gives $\nnuc\simeq\nnux$.}
\begin{equation}
\eta=\frac{\sigma^2}{4M^2g}>1.
\end{equation}
Such a gyroscopic pendulum is known as a ``sleeping top''
because the pendulum ``sleeps'' in the upright position
defying the effect of the gravity 
(see, \textit{e.g.}, Ref.~\cite{Scarborough:1958fk}
for more discussions).

We note that the period of nutation $T_\mathrm{nut}$ of the flavor pendulum
is infinite if $\thetav=\pi/2$. For a symmetric bipolar system
$T_\mathrm{nut}\propto |\ln\qthetav|$ if $\thetav\simeq\pi/2$
(see Sec.~\ref{sec:pendulum-eom} and also Ref.~\cite{Hannestad:2006nj}). 
One expects similarly long nutation periods for asymmetric bipolar systems
in the region where $n_\nu\simeq\nnuc$.
In the same region, $\vartheta_1$ and $\vartheta_2$ change
very quickly in the pure precession limit.
As a result, the condition in Eq.~\eqref{eq:pure-precession-cond} 
is usually violated in realistic environments
and significant nutation can appear for $n_\nu<\nnuc$
[see, \textit{e.g.}, Fig.~\ref{fig:top-theta}(b)].

\subsection{Precession frequency%
\label{sec:omega}}

\begin{figure}
\begin{center}
\includegraphics*[width=\myfigwid, keepaspectratio]{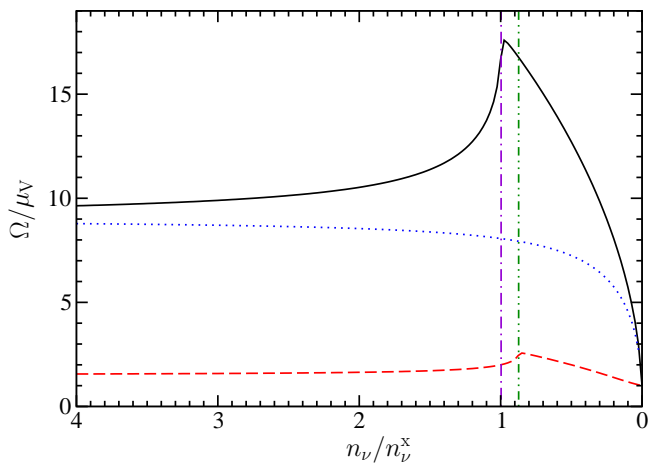}
\end{center}
\caption{\label{fig:omega}(Color online)
Scaled precession frequency $\Omega/\muV{}$ 
of the flavor pendulum as a function of
neutrino number density $n_\nu$. The pendulum is assumed to be in
the pure precession mode. 
The vacuum mixing angle is $\qthetav=0.01$ for the solid
and dashed lines, and $\thetav=0.6$ for the dotted line.
The asymmetry parameter is $\alpha=0.8$ for the solid and
dotted lines and is $0.2$ for the dashed line. The vertical dot-dashed
line and dot-dot-dashed line correspond to $n_\nu=\nnuc$
for $\alpha=0.8$ and $0.2$, respectively.}
\end{figure}

Using Eq.~\eqref{eq:omega} we have calculated the precession
frequency $\Omega$ of the flavor pendulum
for several scenarios assuming the pendulum is always 
in the pure precession mode. The results are plotted in Fig.~\ref{fig:omega}.
The precession frequency $\Omega$ asymptotically approaches the
synchronization frequency $\wsync$ in the synchronized regime
($n_\nu/\nnux\gg1$) as $n_\nu$ becomes larger and larger.
On the other hand, $\Omega$ changes steeply
 with $n_\nu$ in the bipolar regime ($n_\nu/\nnux\lesssim1$).
As $n_\nu$ reaches 0,
$\Omega=\muV{}$ and the flavor pendulum
 precesses with the vacuum oscillation frequency
of the dominant neutrino species ($\nu_e$ in this case).
 
We also note that, for the inverted mass hierarchy scenario with
$\thetav\simeq\pi/2$, the precession frequency $\Omega$ reaches its maximum
at $n_\nu\simeq\nnuc$. Using Eqs.~\eqref{eq:omega} and \eqref{eq:x1-x2-sol},
we obtain
\begin{equation}
\Omega|_{n_\nu=\nnuc}=1+\frac{2\sqrt{\alpha}}{1-\sqrt{\alpha}}
\quad\text{for}\,\thetav=\frac{\pi}{2},
\end{equation} 
which is $17.9$ for $\alpha=0.8$ and $2.6$ for $\alpha=0.2$.
These values  agree well with the solid and dashed lines in 
Fig.~\ref{fig:omega} which assume $\qthetav=0.01$.

\subsection{Equipartition of energies?}

\begin{figure}
\begin{center}
\includegraphics*[width=\myfigwid, keepaspectratio]{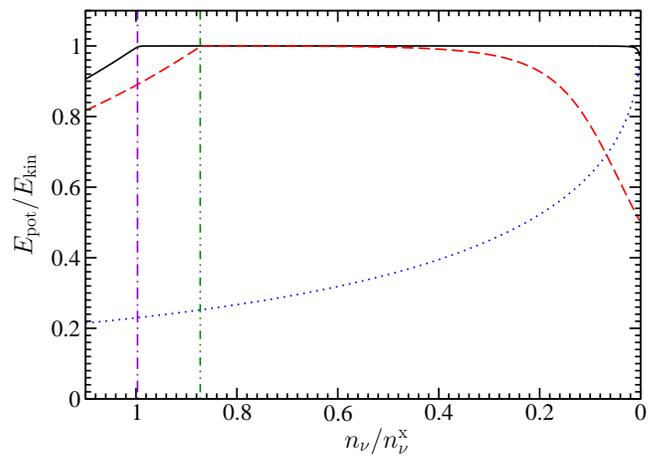}
\end{center}
\caption{\label{fig:kpratio}(Color online)
Ratio of the  potential energy $E_\mathrm{pot}$ of the flavor pendulum
to  its kinetic energy $E_\mathrm{kin}$ [Eq.~\eqref{eq:top-energy}]
as functions of neutrino number density $n_\nu$. 
The parameters and the meaning of the lines
are the same as those in Fig.~\ref{fig:omega}.}
\end{figure}

Ref.~\cite{Hannestad:2006nj} observed 
that, for $\thetav\simeq\pi/2$, the total energy of a
flavor pendulum begins to be approximately equipartitioned between
its  potential and kinetic energies when $n_\nu$ reaches $\nnuc$.
Using Eq.~\eqref{eq:x1-x2-sol} one can show that (see Appendix 
\ref{app:equal-partition}),
for the extreme case $\thetav=\pi/2$, the ratio of energies is
$E_\mathrm{pot}/E_\mathrm{kin}\simeq1$ to $\mathcal{O}(\nnuc/n_\nu-1)$
at $n_\nu\simeq\nnuc$ if the flavor pendulum stays in the pure
precession mode.

Ref.~\cite{Hannestad:2006nj} also mentioned ``an important detail''
that energy equipartition cannot hold all the way down to very small
$n_\nu$ because $E_\mathrm{pot}$ has a finite positive minimum and 
$E_\mathrm{kin}$ ultimately reaches 0. This is illustrated in
Fig.~8 of the same reference. However, we can show that 
(see Appendix \ref{app:equal-partition})
\begin{equation}
\frac{E_{\rm pot}}{E_{\rm kin}}\simeq1-\frac{(1-\alpha)\cos^22\thetav}
{1+\alpha-2\alpha\cos2\thetav}
\quad\text{for}\,\left(\frac{n_\nu}{n_\nu^0}\right)\ll1
\label{eq:kpratio-n0}
\end{equation}
if the flavor pendulum stays in the pure precession mode.

In Fig.~\ref{fig:kpratio} we plot the ratio
$E_\mathrm{pot}/E_\mathrm{kin}$ as a function $n_\nu$
for three different bipolar systems in the pure precession mode
with various choices of $\alpha$ and $\thetav$.
Indeed, for the bipolar systems with $\thetav\simeq\pi/2$,
$E_\mathrm{pot}/E_\mathrm{kin}$ reaches 1 at $n_\nu=\nnuc$
and does not change much for a significant range of
 $n_\nu$. This is especially true for $\alpha\simeq1$.
In the limit $n_\nu=0$, $E_{\rm pot}/E_{\rm kin}\simeq 0.94$ and 0.5 
for $\thetav\simeq\pi/2$ but
$\alpha=0.8$ and 0.2, respectively. In the same limit, 
$E_{\rm pot}/E_{\rm kin}\simeq 0.98$ for $\thetav=0.6$
and $\alpha=0.8$. These results agree well
with Eq.~\eqref{eq:kpratio-n0}.

\section{Neutrino Oscillations in Supernovae%
\label{sec:supernovae}}

Refs.~\cite{Duan:2006an,Duan:2006jv} have presented
two sets of simulations using the ``single-angle''  and
``multi-angle'' approximations, respectively.
These simulations together with the simple analytical
and numerical models discussed in the previous sections
represent approximations to the real supernova neutrino
oscillation problem at three different levels of complexity.

The analytical and numerical calculations performed in
this paper assume that the neutrino gas is homogeneous and isotropic
and is represented by two mono-energetic neutrino and/or antineutrino
species. 

The single-angle simulations increase the complexity
by allowing each neutrino species (4 in the $2\times2$ case)
to have continuous energy distributions. It assumes that the flavor
evolution histories of neutrinos
propagating along different trajectories are the same as those
of neutrinos emitted radially from the neutrino sphere. 
Although the single-angle approximation incorporates the
angle dependence of neutrino-neutrino forward scattering 
into the ``effective neutrino density'' \cite{Duan:2006an},
it  still assumes  that neutrino on 
all trajectories evolve similarly. 
 
The multi-angle simulations are by far the
most sophisticated treatment of the problem. In these calculations
neutrinos and antineutrinos have not  only continuous energy
distributions but also continuous angular distributions.
The most important improvement implemented in the multi-angle
simulations is that the flavor evolution of neutrinos and
antineutrinos (with a wide range of energies) propagating along
different trajectories is followed independently.

In this section we will first apply our simple models to
the single-angle calculations. We will discuss the onset of neutrino
flavor conversion in both the inverted 
and normal mass hierarchy scenarios (Secs.~\ref{sec:start-inverted}
and \ref{sec:start-normal}). We will also investigate the
precession mode of the neutrino gas in supernovae and its effects
(Sec.~\ref{sec:precession-Pnunu}). Finally, we will offer some
new analyses of the multi-angle simulations and comment on
the collectivity of neutrino flavor transformation in supernovae
(Sec.~\ref{sec:collectivity}).

\subsection{Onset of neutrino flavor conversion in the inverted mass hierarchy
scenario%
\label{sec:start-inverted}}

The simulations presented in
Refs.~\cite{Duan:2006an,Duan:2006jv} for
the inverted neutrino mass hierarchy scenario 
all have $\qthetav=0.1$.
According to the discussions in Sec.~\ref{sec:nnuc},
bipolar neutrino systems with vacuum mixing angle $\thetav\simeq\pi/2$
can start flavor conversion after the neutrino number density $n_\nu$
drops below some critical value $\nnuc$. Although the conclusion
was made in the absence of an ordinary matter background, it has been shown that
the evolution of bipolar systems is not changed qualitatively
even in a dominant matter background as long as
$\thetav\simeq\pi/2$ \cite{Duan:2005cp,Hannestad:2006nj}.

For a rough estimate of the radius where
$n_\nu=\nnuc$, we assume that the neutrino gas behaves 
in a way similar to the simple bipolar system initially
consisting of $\nu_e$ and $\bar\nu_e$ with energies
$E_{\nu_e}=11\,\mathrm{MeV}$ and $E_{\bar\nu_e}=16\,\mathrm{MeV}$,
respectively. (These values are the same as
the average energies of $\nu_e$ and $\bar\nu_e$
in the simulations.) In a properly chosen corotating frame,
the evolution of this simplified bipolar system is the same as that of
a gas initial consisting of mono-energetic $\nu_e$ and $\bar\nu_e$
with energy \cite{Duan:2005cp}
\begin{subequations}
\begin{align}
\overline{E}_\nu&\equiv\left[\frac{1}{2}
\left(\frac{1}{E_{\nu_e}}+\frac{1}{E_{\bar\nu_e}}\right)\right]^{-1}\\
&\simeq 13\,\mathrm{MeV}.
\end{align}
\end{subequations}
With the luminosities
of all neutrino species being the same and $L_\nu=10^{51}\,\mathrm{erg/s}$,
the ratio of the number densities of the two neutrino species is
\begin{equation}
\alpha=\frac{L_\nu/E_{\bar\nu_e}}{L_\nu/E_{\nu_e}}\simeq0.69.
\end{equation}
Therefore, the critical neutrino number density is
[see Eq.~\eqref{eq:nnuc}]
\begin{subequations}
\begin{align}
\nnuc&=\frac{\delta m^2}{\sqrt{2}\GF \overline{E}_\nu}(1-\sqrt{\alpha})^{-2}\\
&\simeq 6.24\times10^{28}\,\mathrm{cm}^{-3}
\end{align}
\end{subequations}
for a mass squared difference $\delta m^2=3\times10^{-3}\,\mathrm{eV}^2$.
Using the single-angle approximation we estimate
the effective neutrino number density to be
\footnote{Eq.~\eqref{eq:neff} is similar to Eq.~(40) in Ref.~\cite{Duan:2006an}
except that we here are not computing the net effective neutrino density and,
therefore, ignore the contribution of antineutrinos.}
\begin{subequations}
\label{eq:neff}
\begin{align}
n_\nu^\eff(r) &= \frac{L_\nu}{4\pi R_\nu^2 E_{\nu_e}}
\left[1-\sqrt{1-\left(\frac{R_\nu}{r}\right)^2}\right]^2 \\
&\simeq (1.25\times10^{32}\,\mathrm{cm}^{-3})
\left[1-\sqrt{1-\left(\frac{R_\nu}{r}\right)^2}\right]^2,
\end{align}
\end{subequations}
where $R_\nu=11\,\mathrm{km}$ is the radius of the neutrino
sphere adopted in the simulation. 
Therefore, $n_\nu^\eff\simeq\nnuc$
at radius $r_\mathrm{c}\simeq52\,\mathrm{km}$. 

From panels (c) and (d) of Fig.~8 in 
Ref.~\cite{Duan:2006an} one sees that, in the
single angle simulation, the flavor conversion starts at the
radius  $\rX\simeq 63\,\mathrm{km}$.
At $r\gtrsim\rX$ the $z$-components of
NFIS's $\sB_{\nu_e}$ experience rapid
oscillations which correspond to the nutation mode of
the flavor pendulum.
The estimated value of $r_\mathrm{c}\simeq52\,\mathrm{km}$ and
the observed value of $\rX\simeq 63\,\mathrm{km}$ 
differ by $\sim10\,\mathrm{km}$.
This difference most likely arises
 because at $r\simeq r_\mathrm{c}$ 
the nutation frequency $T_\mathrm{nut}^{-1}$ of the flavor pendulum is very
small. 
Consequently, there is a delay before significant nutation
amplitude can develop. On the other hand, smaller nutation
frequency implies less adiabatic evolution 
[see Eq.~\eqref{eq:pure-precession-cond}]. So once developed,
the nutation amplitude will be large.
The oscillation amplitudes of $\langle s_{\nu_e z}\rangle$
and $\langle s_{\bar\nu_e z}\rangle$ are indeed large
as shown in Fig.~8 of Ref.~\cite{Duan:2006an}.

We note that the region ($r_\mathrm{c}\lesssim r\lesssim\rX$) where 
 the nutation modes are to be excited is roughly the same region
where the chaos-like phenomenon 
shown in Fig.~12 of Ref.~\cite{Duan:2006an} occurs. 
In this region the differences of two almost identical systems
can grow exponentially as they evolve.

\subsection{Onset of neutrino flavor conversion in the normal mass hierarchy
scenario%
\label{sec:start-normal}}

The simulations presented in
Refs.~\cite{Duan:2006an,Duan:2006jv} for
the normal neutrino mass hierarchy scenario 
all have $\thetav=0.1$.
According to the discussions in Sec.~\ref{sec:asymmetric-sys},
a bipolar neutrino system with vacuum mixing angle $\thetav\ll 1$
corresponds to a flavor pendulum that oscillates in a very limited region
near the bottom of the potential well, and therefore, does not experience much
flavor transformation. However, this conclusion
only applies in the absence of an ordinary matter background.

In the presence of a matter background,
it has been shown that,  in the synchronized regime,
neutrinos and antineutrinos of all the species and energies
go through an MSW-like resonance simultaneously in the same way 
as does a neutrino with the characteristic energy 
$E_\mathrm{sync}$ in the conventional MSW picture \cite{Pastor:2002we}.
A similar phenomenon may also occur in bipolar systems in
the bipolar regime (\textit{i.e.}, outside  the synchronized regime)
as suggested in Ref.~\cite{Duan:2005cp}. If this is true, 
the dominant neutrino species are changed from $\nu_e$ and $\bar\nu_e$ 
to $\nu_{\mu,\tau}$ and $\bar\nu_{\mu,\tau}$. 
Using the corotating frame, one can show that
the evolution of a $\bar\nu_{\mu,\tau}$-$\nu_{\mu,\tau}$ gas with the 
normal mass hierarchy is similar to that of a $\nu_e$-$\bar\nu_e$
gas with the inverted mass hierarchy \cite{Duan:2005cp}, and
the flavor pendulum is raised from the bottom position to the top position
because of the change in the dominant neutrino species. 
Bipolar systems can subsequently develop nutation modes after
the collective MSW-like resonance.

From panels (a) and (b) of Fig.~8 in 
Ref.~\cite{Duan:2006an} one sees that, in the
single angle simulation, the $z$-components of
NFIS's $\sB_{\nu_e}$ suddenly change at radius
$\rX\simeq88\,\mathrm{km}$ and oscillate rapidly afterwards.
This corresponds to the initial collective MSW-like resonance
followed by the nutation modes.
We note that the observed value of $\rX$ in the simulation
is larger than the value of $74\,\mathrm{km}$ estimated for
 the fully synchronized limit \cite{Duan:2006an}.
The difference arises partly because the MSW-like resonance
actually occurs in the bipolar regime in this case.

\subsection{Precession mode and final neutrino survival probabilities%
\label{sec:precession-Pnunu}}

As shown in Sec.~\ref{sec:asymmetric-sys} bipolar systems generally
are in both precession and nutation modes. This is indeed seen
in the single-angle simulations for both the normal and inverted mass
hierarchies. In Fig.~8 of Ref.~\cite{Duan:2006an}, 
the $x$- and $y$-components of NFIS's $\sB_{\nu_e}$ 
oscillate with an approximate phase difference of $\pi/2$,
signifying precession in the $x$-$y$ plane.

In Sec.~\ref{sec:pure-precession} we have argued that, 
in the absence of an ordinary matter background,
the intrinsic precession angular velocity
$\bm{\Omega}$ of the bipolar system as a whole should 
be in the same direction as that of a single neutrino of the
dominant species. Therefore, we expect bipolar systems dominated by neutrinos
to tend to precess around $-\HV$. In the presence of a matter
background, bipolar systems will also tend to precess around the
direction opposite to
\begin{equation}
\Hb_e\equiv-\basef{z}\sqrt{2}\GF n_e,
\end{equation}
where $n_e$ is the net electron number density. For the inverted mass
hierarchy with $\thetav\simeq\pi/2$ and $\HV\simeq-\basef{z}$, 
the intrinsic  $\bm{\Omega}$ of the flavor pendulum
is roughly in the same direction as the precession stemming from
the matter background. In this case, the combined precession does
not change direction. As a result,
the precession of NFIS's is always roughly around
$-\HV\simeq\basef{z}$ for the inverted mass hierarchy. 

For the normal mass hierarchy with
$\thetav\ll 1$ and $\HV\simeq\basef{z}$,  the 
intrinsic  $\bm{\Omega}$ of the flavor pendulum
is roughly in the opposite direction to that
of the precession due to
the matter background, and the combined precession may change its direction.
However, it is expected that the matter background becomes
negligible after the collective MSW-like resonance.
Therefore, the NFIS's precess roughly in
the direction of $-\HV\simeq-\basef{z}$ in the region $r\gtrsim\rX$
for the normal mass hierarchy.
According to Fig.~8 of Ref.~\cite{Duan:2006an} NFIS's
indeed precess around $\basef{z}$ for the inverted mass
hierarchy scenario and  $-\basef{z}$ for the normal mass
hierarchy scenario.

Ref.~\cite{Duan:2006an} has shown that (see Fig.~9 of that reference), 
for the inverted mass hierarchy scenario and at large radius,
neutrinos with energies below $E_\mathrm{C}\simeq 9\,\mathrm{MeV}$
are mostly in their initial flavors while neutrinos with larger energies
and most antineutrinos can be completely converted to other flavors
in the limit of large neutrino luminosity $L_\nu$.
For the normal mass hierarchy scenario, 
neutrinos with energies below $E_\mathrm{C}\simeq 9\,\mathrm{MeV}$
are almost completely converted to other flavors
while  neutrinos with larger energies
and nearly all antineutrinos are mostly in their initial flavors
in the limit of large $L_\nu$. 
Ref.~\cite{Duan:2006an} suggested that this phenomenon is
related to the precession of NFIS's when
neutrino number densities decrease and the bipolar configuration
starts to break down (see Fig.~10 of that reference).

We note that the precession of NFIS's due to the matter 
background $\Hb_e$ is the same for all neutrinos, and we can essentially
ignore it in a reference frame $\mathcal{F}_1$ 
rotating with angular velocity $-\Hb_e$.
In this corotating frame $\mathcal{F}_1$, the e.o.m.~of NFIS $\sB_i$ is
\begin{subequations}
\begin{align}
\frac{\ud}{\ud t}\sB_i &= \sB_i\times\Hb_\mathrm{tot}\\
&\equiv\sB_i\times(\muV{,i}\HV+\Hb_\nu)\\
&\equiv\sB_i\times(\muV{,i}\HV+\mu_\nu\sum_j n_{\nu,j}\sB_j)
\end{align}
\end{subequations}
where $\Hb_\nu$ is the effective ``magnetic field'' generated by 
all other NFIS's. We assume that all NFIS's and
$\Hb_\nu$ rotate with a constant angular velocity 
$-\Omega\HV$. The problem becomes very simple
in a reference frame $\mathcal{F}_2$ which rotates
relative to $\mathcal{F}_1$ with angular velocity $-\Omega\HV$. 
In $\mathcal{F}_2$
\begin{subequations}
\begin{align}
\frac{\ud}{\ud t}\sB_i &= \sB_i\times\widetilde{\Hb}_\mathrm{tot}\\
&\equiv\sB_i\times[(\muV{,i}-\Omega)\HV+\widetilde{\Hb}_\nu],
\end{align}
\end{subequations}
where both $\widetilde{\Hb}_\mathrm{tot}$ and
$\widetilde{\Hb}_\nu$ are not rotating
\footnote{We have ignored the rotation of $\HV$ in the corotating
frames $\mathcal{F}_1$ and $\mathcal{F}_2$ because $\HV$ is roughly
in the same or opposite direction as the rotation axis
 for  $\thetav\ll1$ or $\thetav\simeq\pi/2$.}.

We first look at a NFIS $\sB_i$ corresponding 
to a neutrino which is initially pure $\nu_e$ at
the neutrino sphere.
Because $\mu_\nu\equiv-2\sqrt{2}\GF$ is negative and the neutrino gas
is initially dominated by $\nu_e$, the NFIS $\sB_i$
must be roughly antialigned with 
$\widetilde{\Hb}_\mathrm{tot}\simeq\widetilde{\Hb}_\nu$ when
neutrino number densities are large. As neutrino number densities
decrease to 0, $\widetilde{\Hb}_\mathrm{tot}\rightarrow(\muV{,i}-\Omega)\HV$. 
If neutrino
number densities decrease so slowly that the process is adiabatic,
$\sB_{i}$ will stay antialigned with $\widetilde{\Hb}_\mathrm{tot}$.
At $n_\nu=0$ the NFIS $\sB_{i}$ can be either aligned
or antialigned with $\HV$ depending on whether $\muV{,i}$
is smaller or larger than $\Omega$. For the inverted mass hierarchy
scenario with $\thetav\simeq\pi/2$ and $\HV\simeq-\basef{z}$,
$\sB_i$ is roughly aligned with $\basef{z}\simeq-\HV$ if
$\muV{}\equiv \delta m^2/(2E_\nu)>\Omega$ and antialigned with
$\basef{z}$ otherwise. Accordingly, at large radii 
neutrinos starting as $\nu_e$ 
are still mostly in the $\nu_e$ flavor if their energies are below
\begin{equation}
\EC\equiv\frac{\delta m^2}{2 \Omega}
\end{equation}
and are almost completely converted to other flavors otherwise.
In  words, one has
\begin{equation}
P_{\nu\nu}(E_\nu)\simeq\left\{\begin{array}{ll}
1&\text{if }E_\nu<\EC,\\
0&\text{if }E_\nu>\EC.
\end{array}\right.
\end{equation}
One can estimate the final neutrino survival probabilities
for other cases in a 
similar fashion. We have summarized the results 
for the relevant scenarios in Table \ref{tab:Pnunu}.

\begin{table}
\caption{\label{tab:Pnunu}Final neutrino survival
probabilities in the adiabatic limit for large $L_\nu$
in both the normal and inverted neutrino mass hierarchy scenarios.} 
\begin{ruledtabular}
\begin{tabular}{c|cc}
&$\thetav\ll 1$ (normal) & $\thetav\simeq\pi/2$ (inverted) \\
\hline
$P_{\nu\nu}(E_\nu<\EC)$ & 0 & 1 \\
$P_{\nu\nu}(E_\nu>\EC)$ & 1 & 0 \\
$P_{\bar\nu\bar\nu}(E_{\bar\nu})$ & 1 & 0 
\end{tabular} 
\end{ruledtabular} 
\end{table}

In this analysis
 we have assumed $\Omega$ to be constant.
This analysis is expected to hold as long as the process is
more or less adiabatic and neutrino number densities decrease slowly.
The predictions from this simple analysis generally agree 
with the results of single-angle numerical simulations presented in
Fig.~9 of Ref.~\cite{Duan:2006an}. The agreement is especially good
for large neutrino luminosities
and in the neutrino sector for which $P_{\nu\nu}(E_\nu)$ has a relatively
sharp transition or jump at $E_\nu\simeq\EC$. This pattern
can be taken as a hallmark
of collective neutrino flavor transformation
because it results from a neutrino background that is in a collective
precession mode.

\subsection{Collectivity and non-collectivity of neutrino oscillations
in supernovae%
\label{sec:collectivity}}

\begin{figure*}
\begin{center}
$\begin{array}{@{}c@{\hspace{\myfigsep}}c@{}}
\includegraphics*[width=\myfigwid, keepaspectratio]{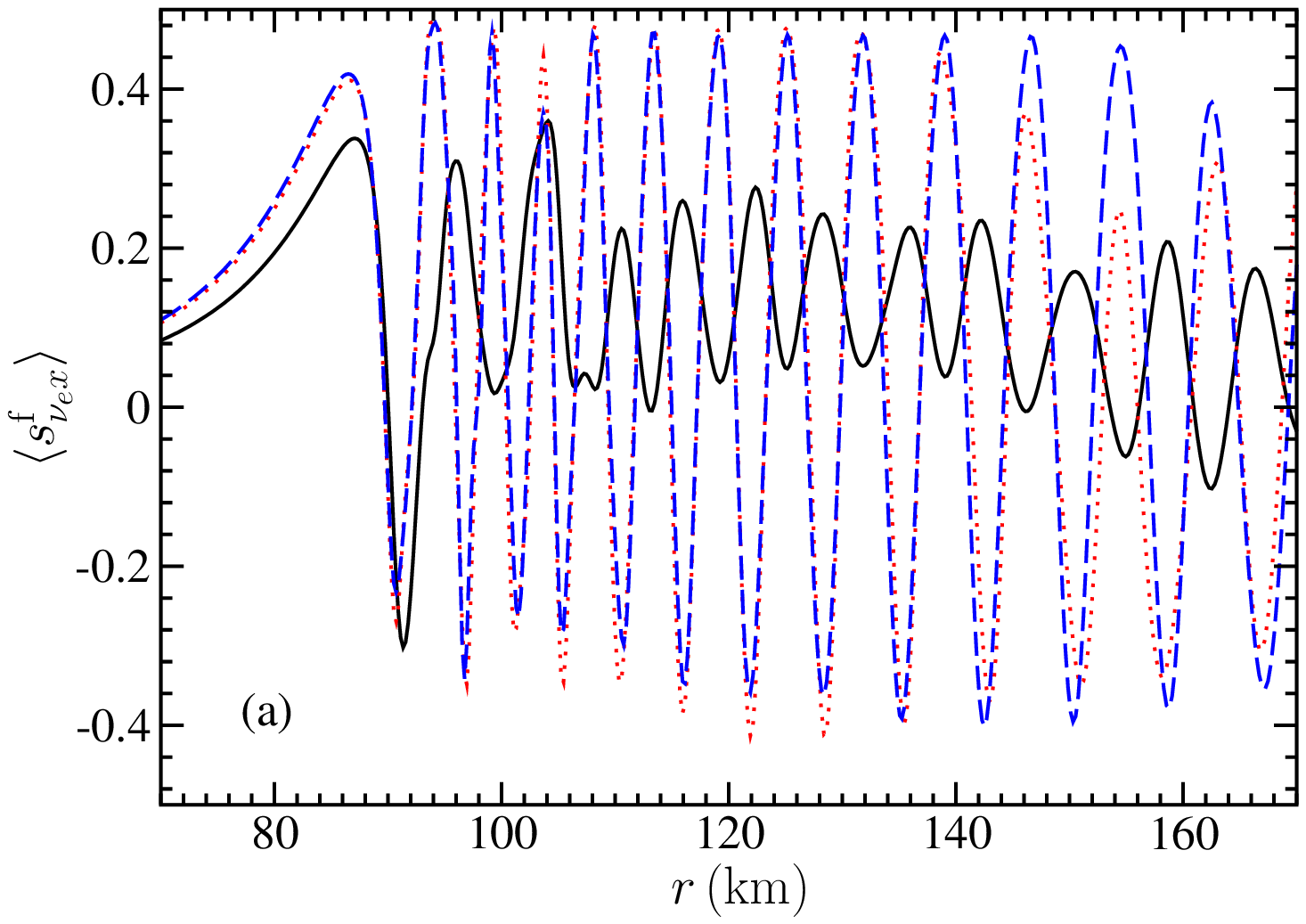} &
\includegraphics*[width=\myfigwid, keepaspectratio]{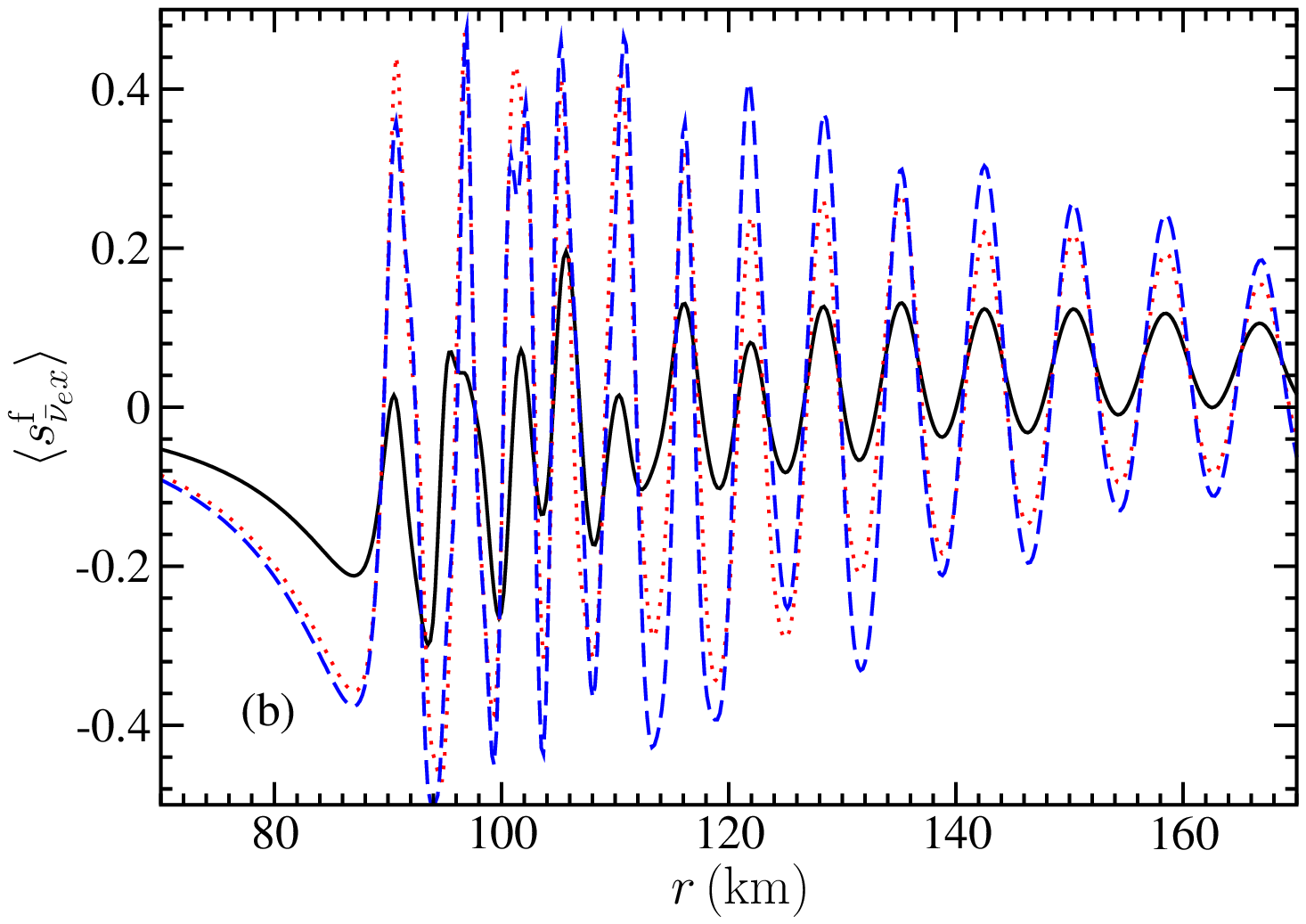} \\
\includegraphics*[width=\myfigwid, keepaspectratio]{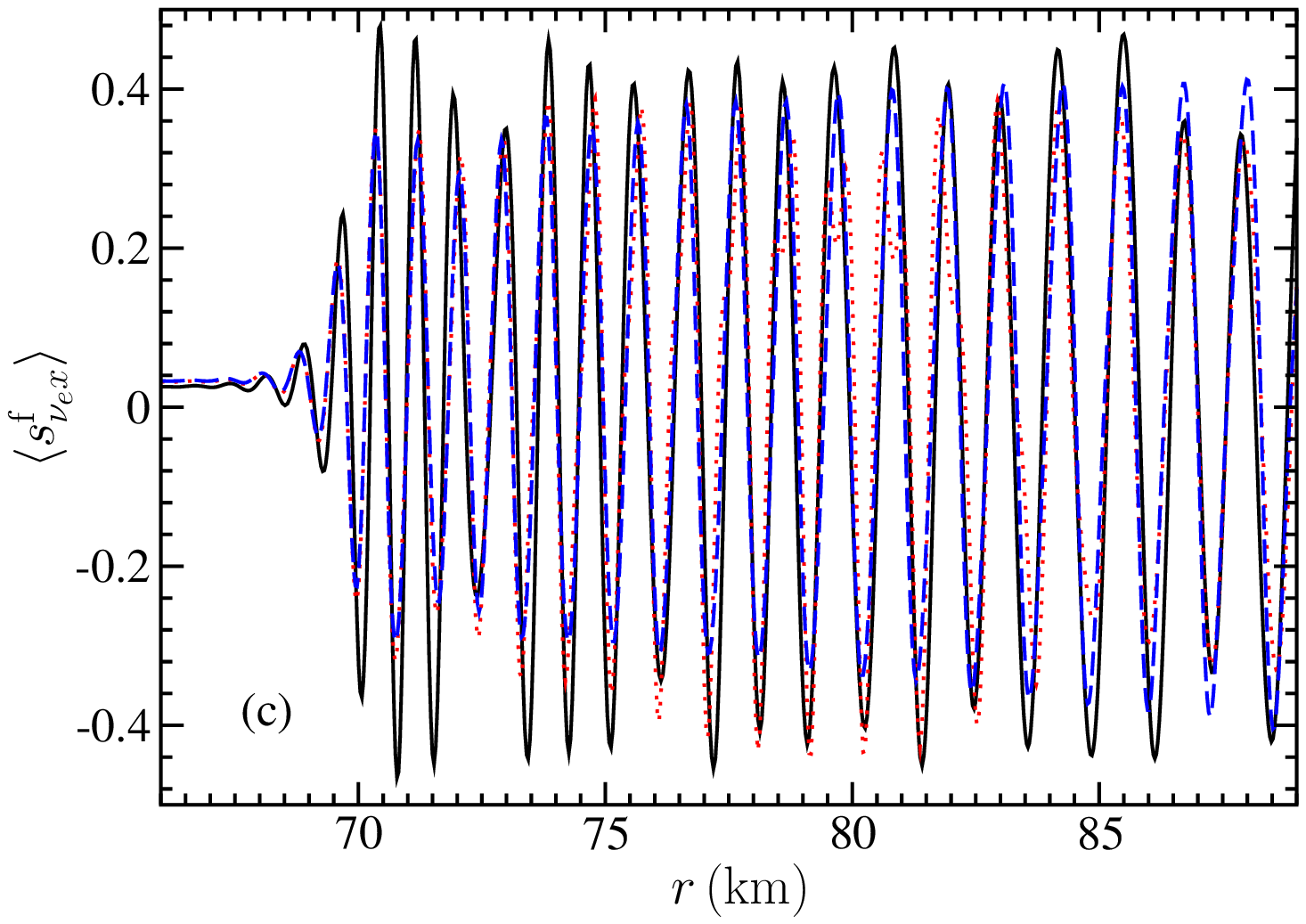} &
\includegraphics*[width=\myfigwid, keepaspectratio]{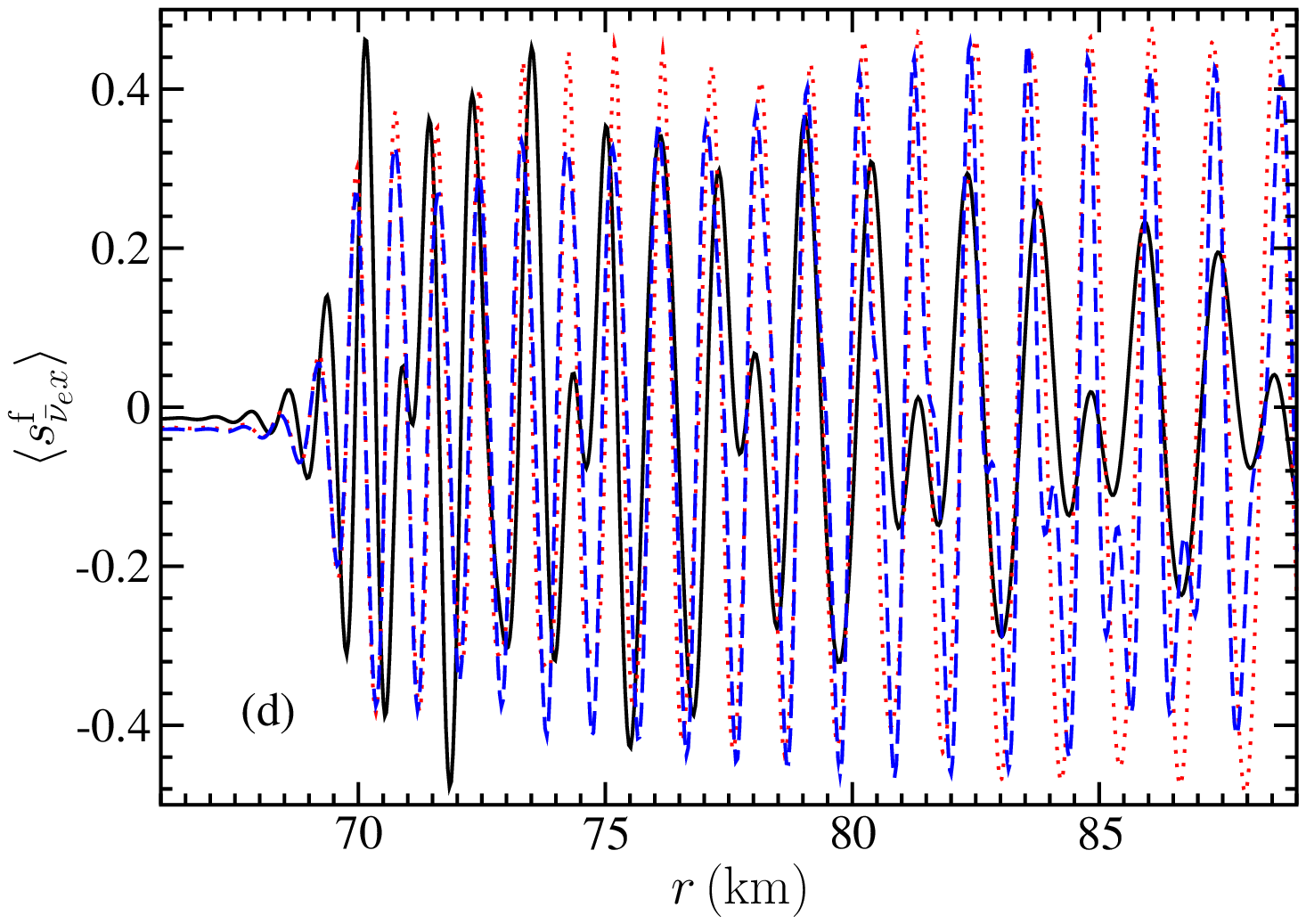}
\end{array}$
\end{center}
\caption{\label{fig:sx-r}(Color online)
The evolution of $x$-components of the average NFIS's of supernova neutrinos
in the flavor basis obtained from the multi-angle simulations as
presented in Fig.~4 of Ref.~\cite{Duan:2006an}.
The left panels are for neutrinos emitted as $\nu_e$, and 
the right panels are for antineutrinos emitted as $\bar\nu_e$.
The top panels employ the normal mass hierarchy with $\thetav=0.1$, and
the bottom panels employ the inverted mass hierarchy with $\qthetav=0.1$.
The solid, dotted and dashed  lines give $\langle s_{\nu x}^\mathrm{f}\rangle$
along the trajectories with $\cos\Theta_0=1$, $0.5$ and $0$, respectively,
where $\Theta_0$ is the emission angle defined in Ref.~\cite{Duan:2006an}.
The neutrino mass-squared difference is taken to be
$\delta m^2=3\times10^{-3}\,\mathrm{eV}^2$.}
\end{figure*}

Single-angle simulations assumed that neutrinos
of the same species and energy all evolve in the same way
even if they are emitted
in different angles from the neutrino sphere. This is not
necessarily always a good approximation as neutrino-neutrino
forward scattering is angle dependent and the 
neutrino density distributions in the supernova environment
are inhomogeneous and anisotropic. Even if neutrinos moving along various
trajectories all have the same flavor content at a given radius,
they produce different  refractive indices  for neutrinos
propagating in different directions. Therefore, 
neutrinos propagating in different directions 
are expected to have different flavor evolution histories.
On the other hand, neutrinos propagating along different trajectories
are coupled to each other through neutrino-neutrino forward scattering.
The correlations among different neutrino trajectories are especially
strong when neutrino fluxes are large. The inhomogeneity/anisotropy
of the environment and the correlation among different neutrino
trajectories act as two opposite ``forces'' which try to break and uphold,
respectively, the collective aspect of neutrino oscillations in supernovae.
At the moment it is difficult to perform an analytical study that can clearly
predict which force wins. Our limited goal here is
to gain insight into the issue of collectivity
of neutrino flavor transformation by analyzing
the multi-angle simulations presented in Refs.~\cite{Duan:2006an,Duan:2006jv}.

As shown in Fig.~2 of Ref.~\cite{Duan:2006jv},
the flavor evolution of neutrinos on each trajectory in multi-angle
simulations looks qualitatively similar to that in the
corresponding single-simulations. However, the oscillations
in neutrino survival probabilities
$P_{\nu\nu}$ and $P_{\bar\nu\bar\nu}$ have different frequencies
for different trajectories.
For vacuum angle $\thetav\ll1$ or 
$\thetav\simeq\pi/2$, the oscillations in $P_{\nu\nu}$ and $P_{\bar\nu\bar\nu}$
represent the nutation of the flavor pendulum. Therefore, the
nutation modes of neutrinos propagating along different trajectories
cannot be viewed as collective.
Indeed, it has recently been shown that
the nutation modes  for symmetric bipolar systems
can quickly  develop large phase differences and ``de-cohere''
 for neutrinos propagating in 
different directions \cite{Raffelt:2007yz}. 

On the other hand, Fig.~3 of Ref.~\cite{Duan:2006jv} shows
that $P_{\nu\nu}(E_\nu)$ obtained by multi-angle simulations 
has a sharp transition at $E_\nu=\EC$
as in single-angle simulations.
In addition, the value of $\EC$ is approximately independent of neutrino
trajectory direction. If this transition  is related to
the precession mode of neutrinos as suggested by Ref.~\cite{Duan:2006an}
and further explained here,
NFIS's corresponding to neutrinos propagating
in different directions must precess with the same frequency.
In Fig.~\ref{fig:sx-r} we have plotted $\langle s_{\nu x}^\mathrm{f}\rangle$,
the $x$-component of the average NFIS's in the flavor basis, as functions of
radius $r$ for three representative trajectories
obtained from the multi-angle simulations 
of Refs.~\cite{Duan:2006an,Duan:2006jv}.
One indeed observes that the NFIS's along various trajectories
are approximately in a single collective precession mode and precess around
$\pm\basef{z}$ with approximately the same frequency.

\section{Conclusions\label{sec:conclusion}}

We have investigated the simple symmetric bipolar system
using the flavor pendulum analogy. We have shown that
an adiabatic invariant of the pendulum motion
can be used to study the evolution of such a bipolar
system when neutrino number densities change slowly with time.
We have also studied an asymmetric bipolar system
using the gyroscopic pendulum analogy. As a gyroscopic pendulum,
a bipolar system generally can be in both the
precession and nutation modes simultaneously
except in the synchronized regime where
only precession is possible.

We have shown that an asymmetric bipolar system
can stay mostly in a pure precession mode as it
transitions from the synchronized regime
into the bipolar regime 
if neutrino number densities decrease slowly.
The precession frequency of the system generally
varies with the neutrino number density and
approaches the synchronization frequency
in the synchronized regime.
For the inverted mass hierarchy case with mixing
angle $\thetav\simeq\pi/2$, we have calculated 
a more accurate value of the critical neutrino number density
below which bipolar systems
can start flavor transformation. Because
supernova neutrinos naturally form asymmetric bipolar systems,
these analyses could be useful for understanding
the qualitative features of neutrino oscillations in supernovae.

We have further analyzed the recent numerical simulations 
of neutrino oscillations in supernovae. 
 These large-scale simulations suggest that neutrinos
traveling on intersecting trajectories
and experiencing destructive quantum interference
nevertheless can be in the collective precession
mode. This mode can result in
sharp transitions in the final
energy-dependent neutrino survival probabilities $P_{\nu\nu}(E_\nu)$ across
all trajectories. This sharp transition in  $P_{\nu\nu}(E_\nu)$
can be taken as a hallmark of collective neutrino flavor transformation.
Moreover, this transition occurs differently for the
normal and inverted neutrino mass hierarchies. Based on this
difference, the neutrino signals from a future galactic supernova
potentially can  be used to determine the actual
neutrino mass hierarchy.

\appendix*
\section{Potential and kinetic energies of 
an asymmetric flavor pendulum%
\label{app:equal-partition}}

Let us compare the kinetic energy $E_\mathrm{kin}$
of the flavor pendulum with its potential energy $E_\mathrm{pot}$
[see Eq.~\eqref{eq:top-energy}] in the pure precession mode. 
The potential energy is defined as 
\begin{equation}
E_{\rm pot}\equiv Mg(1-\HV\cdot\rb)=\muV{}(q-\qb\cdot\HV),
\end{equation}
where 
\begin{equation}
\qb\equiv\frac{\Qb}{n_\nu}=\sB_1-\alpha\sB_2+\frac{n_\nu^0}{n_\nu}\HV.
\end{equation}
Its kinetic energy is defined as
\begin{equation}
E_{\rm kin}\equiv\frac{\Jb^2}{2M}=\frac{\muV{}}{2}
\left(\frac{n_\nu}{n_\nu^0}\right)\Jb^2,
\end{equation}
where
\begin{equation}
\Jb\equiv\sB_1+\alpha\sB_2.
\end{equation}
According to Eq.~\eqref{eq:x1-x2-sol}, we have
$\sin^2\vartheta_1\simeq\alpha \sin^2\vartheta_2$ 
and $\sin^2\vartheta_2\simeq 2\delta/\sqrt{\alpha}$
if
\begin{equation}
\delta\equiv\frac{\nnuc}{n_\nu}-1\ll 1.
\end{equation}
It is straightforward to show that
\begin{widetext}
\begin{subequations}
\begin{align}
q&=\sqrt{(\sB_1-\alpha\sB_2)^2+2\left(\frac{n_\nu^0}{n_\nu}\right)
(\sB_1\cdot\HV-\alpha\sB_2\cdot\HV)+\left(\frac{n_\nu^0}{n_\nu}\right)^2}\\
&=\sqrt{\frac{1+\alpha^2-2\alpha\cos(\vartheta_1+\vartheta_2)}{4}+
\left(\frac{n_\nu^0}{n_\nu}\right)(\cos\vartheta_1-\alpha\cos\vartheta_2)
+\left(\frac{n_\nu^0}{n_\nu}\right)^2}\\
&=\frac{1+\alpha}{2}-\frac{n_\nu^0}{n_\nu}+\mathcal{O}(\delta^2).
\label{eq:q-lim1}
\end{align}
\end{subequations}
\end{widetext}
In addition, we have
\begin{subequations}
\begin{align}
\qb\cdot\HV&=\sB_1\cdot\HV-\alpha\sB_2\cdot\HV+\frac{n_\nu^0}{n_\nu}\\
&\simeq-\frac{1+\alpha}{2}+\frac{n_\nu^0}{n_\nu}+\sqrt{\alpha}\delta.
\label{eq:qz-lim1}
\end{align}
\end{subequations}
Combining Eqs.~\eqref{eq:q-lim1} and \eqref{eq:qz-lim1}, we obtain
\begin{subequations}
\begin{align}
E_{\rm pot}&\simeq\muV{}\left[1+\alpha-2\left(\frac{n_\nu^0}{n_\nu}\right)
-\sqrt{\alpha}\delta\right]\\
&=\muV{}\left[\frac{(1+\sqrt{\alpha})^2}{2}-
\left(\frac{1+\alpha}{2}\right)\delta\right].
\end{align}
\end{subequations}

Similarly, we have
\begin{subequations}
\begin{align}
\Jb^2&=\frac{1+\alpha^2+2\alpha\cos(\vartheta_1+\vartheta_2)}{4}\\
&\simeq\left(\frac{1-\alpha}{2}\right)^2+
\frac{\sqrt{\alpha}(1-\sqrt{\alpha})^2}{2}\delta,
\end{align}
\end{subequations}
and
\begin{subequations}
\begin{align}
E_{\rm kin}&\simeq\frac{\muV{}}{1+\delta}
\left[\frac{(1+\sqrt{\alpha})^2}{2}+\sqrt{\alpha}\delta\right]\\
&\simeq\muV{}\left[\frac{(1+\sqrt{\alpha})^2}{2}-
\left(\frac{1+\alpha}{2}\right)\delta\right].
\end{align}
\end{subequations}
Therefore, we obtain $E_{\rm pot}/E_{\rm kin}\simeq 1$ to 
${\mathcal{O}}(\delta)$
in the limit $\delta\ll 1$.

One can also estimate the   potential
and kinetic energies of the flavor pendulum 
in the limit $n_\nu/n_\nu^0\ll 1$. In this limit we have
$\sB_1\cdot\HV\simeq\cos\vartheta_1/2$, $\sB_2\cdot\HV\simeq -1/2$, and
$\sB_1\cdot\sB_2\simeq -\cos\vartheta_1/4$. Therefore,
\begin{widetext}
\begin{subequations}
\begin{align}
q&=\sqrt{(\sB_1-\alpha\sB_2)^2+2\left(\frac{n_\nu^0}{n_\nu}\right)
(\sB_1\cdot\HV-\alpha\sB_2\cdot\HV)+\left(\frac{n_\nu^0}{n_\nu}\right)^2}\\
&=\left(\frac{n_\nu^0}{n_\nu}\right)\sqrt{1+2\left(\frac{n_\nu}{n_\nu^0}\right)
(\sB_1\cdot\HV-\alpha\sB_2\cdot\HV)+\left(\frac{n_\nu}{n_\nu^0}\right)^2
(\sB_1-\alpha\sB_2)^2}\\
&\simeq\frac{n_\nu^0}{n_\nu}+\sB_1\cdot\HV-\alpha\sB_2\cdot\HV
+\frac{1}{2}\left(\frac{n_\nu}{n_\nu^0}\right)[(\sB_1-\alpha\sB_2)^2-
(\sB_1\cdot\HV-\alpha\sB_2\cdot\HV)^2]\\
&\simeq\qb\cdot\HV+\left(\frac{n_\nu}{n_\nu^0}\right)\frac{\sin^2\vartheta_1}{8},
\end{align}
\end{subequations}
\end{widetext}
which gives
\begin{equation}
E_{\rm pot}\simeq\frac{\muV{}}{8}\left(\frac{n_\nu}{n_\nu^0}\right)\sin^2\vartheta_1.
\label{eq:Epot-lim2}
\end{equation}
Similary, we have
\begin{subequations}
\begin{align}
E_{\rm kin}&=\frac{\muV{}}{2}\left(\frac{n_\nu}{n_\nu^0}\right)
(\sB_1+\alpha\sB_2)^2\\
&\simeq\frac{\muV{}}{8}\left(\frac{n_\nu}{n_\nu^0}\right)
(1+\alpha^2-2\alpha\cos\vartheta_1).
\label{eq:Ekin-lim2}
\end{align}
\end{subequations}
Combining Eqs.~\eqref{eq:Epot-lim2} and \eqref{eq:Ekin-lim2}, we obtain
\begin{equation}
\frac{E_{\rm pot}}{E_{\rm kin}}\simeq\frac{\sin^2\vartheta_1}
{1+\alpha^2-2\alpha\cos\vartheta_1}.
\end{equation}
In the limit $n_\nu/n_\nu^0\ll1$, $\cos\vartheta_1\simeq(1-\alpha)
\cos2\thetav+\alpha$, and the above equation reduces to
\begin{equation}
\frac{E_{\rm pot}}{E_{\rm kin}}\simeq1-\frac{(1-\alpha)\cos^22\thetav}
{1+\alpha-2\alpha\cos2\thetav}.
\end{equation}

\begin{acknowledgments}
This work was supported in part by 
NSF grant PHY-04-00359,
the TSI collaboration's DOE SciDAC grant at UCSD, and
DOE grant DE-FG02-87ER40328 at UMN.
This work was also supported in part by the LDRD Program
and Open Supercomputing at LANL, and by
the National Energy Research Scientific Computing Center through
the TSI collaboration using Bassi, and the San Diego Supercomputer 
Center through the Academic Associates Program using DataStar. 
We would like to thank A.~Friedland,  T.~Goldman, 
J.~Hidaka and M.~Patel
for valuable conversations.
\end{acknowledgments}

\bibliography{ref}

\end{document}